\documentclass[preprint2]{aastex}
\usepackage{color}

\newcommand{\amat}{\arrfont{C}}                 
\newcommand{\arrfont}[1]{\mbox{\sffamily$\textbf{#1}$}}
\newcommand{\be}{\begin{equation}}
\newcommand{\ci}{{\rm MICI}}                    
\newcommand{\conlab}[1]{#1_{\rm{c}}}            
\newcommand{\cpress}{p}                         
\newcommand{\diag}{{\rm diag}} 
\newcommand{\Dmat}{\arrfont{D}}
\newcommand{\DTmat}{\arrfont{D}^{\rm T}}
\newcommand{\ee}{\end{equation}}
\newcommand{\eq}[1]{(\ref{#1})}
\newcommand{\eqr}{{\cal R}_{\rm eq}}
\newcommand{\gl}{{\rm G}}
\newcommand{\gll}{{\rm GL}}

\newcommand{\gnum}{\lavec{g}}                   

\newcommand{\Jmat}{\mathbf{\Phi}}             
\newcommand{\lavec}[1]{{\mathbf{#1}}}         
\newcommand{\Lmat}{\arrfont{L}}                 
\newcommand{\Ltwo}{{\cal L}_2}                   
\newcommand{\mask}{\mathbf{\mathsf{\Pi}}}   
\newcommand{\Mmat}{\arrfont{M}}                 
\newcommand{\pdomain}{\set{D}}                  
\newcommand{\pnum}{\lavec{p}}                   
\newcommand{\ppmnum}{\lavec{{p}}^{\pm}} 
\newcommand{\polynomialsset}[1]{\set{P}_{#1}}   

\newcommand{\Qmat}{\arrfont{A}}                 
\newcommand{\Rn}{{\sf {R_v}}}
\newcommand{\Sec}[1]{section \ref{#1}}
\newcommand{\set}[1]{\mathbf{#1}}               

\newcommand{\znum}{\lavec{Z}}                   
\newcommand{\zpnum}{\lavec{Z}^{+}}

\newcommand{\zmnum}{\lavec{Z}^{-}}

\newcommand{\zpmnum}{\lavec{Z}^{\pm}}
\newcommand{\zmpnum}{\lavec{Z}^{\mp}}

\def\eg{{\it e.g.}\ }

\def\al{Alfv\'en\ }

 \def\B{{\lavec{B}}} \def\j{{\lavec{j}}} \def\b{{\lavec{b}}} \def\u{{\lavec{u}}}   \def\Z{{\lavec{Z}}}  

\shorttitle{A comparison of spectral element and finite difference methods}
\shortauthors{Ng et al.}

\begin{document}

\title{A comparison of spectral element and finite difference methods using statically refined 
nonconforming grids for the MHD island coalescence instability problem}
 
\author{ C. S. Ng\altaffilmark{1}, D. Rosenberg\altaffilmark{2}, K. Germaschewski\altaffilmark{1}, 
         A. Pouquet\altaffilmark{2}, and A. Bhattacharjee\altaffilmark{1}}
\altaffiltext{1}{ Space Science Center, University of New Hampshire, 39 College Road, Durham, NH 03824, USA}
\altaffiltext{2}{TNT/IMAGe, National Center for Atmospheric Research, P.O. Box 3000, Boulder, CO 80307-3000, USA}

\begin{abstract}
A recently developed spectral-element adaptive refinement incompressible magnetohydrodynamic (MHD) code 
[Rosenberg, Fournier, Fischer, Pouquet, J. Comp. Phys. 215, 59-80 (2006)] 
is applied to simulate the problem of MHD island coalescence instability (\ci) in two dimensions. 
\ci\ is a fundamental MHD process that can produce sharp current layers
and subsequent reconnection and heating in a high-Lundquist number plasma
such as the solar corona [Ng and Bhattacharjee, Phys. Plasmas, 5, 4028 (1998)].  
Due to the formation of thin current layers, 
it is highly desirable to use adaptively or statically refined grids to resolve them, 
and to maintain accuracy at the same time.  
The output of the spectral-element static adaptive refinement simulations are compared with simulations 
using a finite difference method on the same refinement grids, 
and both methods are compared to  pseudo-spectral simulations with uniform grids as baselines.
It is shown that with the statically refined grids roughly scaling linearly with effective resolution,
spectral element runs can maintain accuracy significantly higher
than that of the finite difference runs, in some cases achieving close to full spectral accuracy.
\\
\end{abstract}

\maketitle

\section{Introduction}
\label{Intro}

In many hydrodynamic or magnetohydrodynamic  (MHD) applications in astrophysics or space physics,
it is essential that a numerical simulation resolve the development of sharp spatial structures accurately.  
While pseudo-spectral methods generally can maintain high accuracy, 
they are mainly applied on more regular geometry and require more uniform grids,
which can make it difficult to reach high resolution in order to resolve sharp isolated structures
especially in flows
dominated by such structures. Static or adaptive mesh refinement (AMR) methods can put more grid points 
in and around isolated structures in order to resolve them, 
but may not achieve similar high-accuracy if low order spatial discretizations are used \cite{rosenberg2007}.  
They are particularly useful in bounded flows, where pseudo--spectral methods are often not optimized.
Therefore, it is of great interest to develop numerical schemes that can combine high accuracy 
and high spatial resolution.  
Adaptive spectral element methods have the potential to do just that, 
providing spectral-like accuracy that can be applied efficiently to resolve isolated structures. 
In this paper, we concentrate on comparing the accuracy of simulation results from a spectral element based AMR code (SE)\cite{rosenberg2006, rosenberg2007} 
and a finite difference based AMR code (FD) \cite{astronum2006,b2005} 
on an astrophysical problem that requires high spatial resolution as well as high accuracy, 
the so--called  MHD island coalescence instability (\ci) problem \cite{nb1998}.  
In order to make meaningful comparisons, we let each code run on essentially the  same non-uniform grid 
(i.e. with the total degrees of freedom in the problem fixed) that is refined a priori in regions of the grid
where the current sheets will form in the \ci.
A separate pseudo-spectral code (PS), running on uniform grids, 
is also used to provide a baseline for the comparisons.  

There is little in the literature about attempts to compare FD and high order methods.
Often, it is simply accepted that the fixed lower order FD methods will be less accurate, but that,
due to their simplicity and efficiency, {\it h}--type grid refinement can always be carried out to 
improve the accuracy for any real problem. We do not attempt in this work to
compare performance metrics of the methods, preferring instead to focus on their ability
to produce accurate solutions. To an extent, then, it is clear that {\it h}-refinement
will improve solution accuracy for many problems; however, previous
work \cite{rosenberg2007} suggests that local high order may be {\it required} in certain
instances. We consider this issue of {\it h}-refinement in FD briefly in this work.
Previous {\it comparative} work in one dimension \cite{basdevant86} demonstrates that spectral methods, 
including SE, are more accurate than FD methods due primarily to dispersion problems in the latter.
Nevertheless, this work concludes that the SE method is not well suited to
the calculation of thin internal layers (structures), particularly when their location is
unknown. This assertion is made because, while accuracy is found to be good, Basdevant et al. 1986 find that 
polynomial orders are required to
be inordinately high even in one dimension. This conclusion is refuted in later work
\cite{mavriplis94}, which found that the SE method is indeed well suited to this type of 
problem provided adaptivity is used (see also \cite{rosenberg2006}). More recent work \cite{st-cyr2007}
also demonstrates that the SE method exhibits smaller errors than a low-order finite volume 
scheme at comparable resolutions for nearly all of a series of tests of a shallow water model on 
a cubed--sphere grid that is either adaptively or statically refined.
However, to the best of our knowledge, ours is the first work that performs a quantitative
comparison of FD with SE methods in divergence--free nonlinear flows in two space dimensions 
with nonconforming (statically) adaptive grids.

We continue our introductory remarks on the \ci\ in \Sec{sec_island_instability}, where we
present details of the problem, and provide a motivation for our work. In the process,
we discuss important properties of the \ci\ as well as some aspects of MHD flows
that will serve our later discussion. We provide details about the simulation set up as 
well as some diagnostic measures used in the comparisons in \Sec{sec_props}.
In \Sec{sec_methods} we present the numerical methods used in sufficient detail to elucidate the results. 
Our results are presented in \Sec{sec_results}, where we offer several types of
accuracy comparisons relevant to MHD flows, and the \ci, in particular.
We conclude in \Sec{sec_conclusion} with a summary of our findings, 
and some discussion about the relative advantages of the methods.

\section{MHD Island Coalescence Instability}
\label{sec_island_instability}
In this section, we provide a brief background for the \ci\ problem 
that motivates us to perform our comparative study in this particular simulation configuration.  

It is well known that a substantial part of our universe is composed of systems of 
plasmas, ionized gases, and conducting fluids.  
Magnetic fields, both fluctuating and large scale, 
play an important role in the physics of these systems.  
In fact, large-scale magnetic fields have long been observed to exist in the solar corona 
(stellar coronae, see e.g. \cite{parker1979} and references therein),
interstellar space (within a galaxy,  see e.g., \cite{forman1985,grimes2005}),
galaxy clusters (see e.g.,  \cite{kellogg1973,sarazin1986}).  
We consider the representation of magnetohydrodynamics (MHD), 
as a starting point of discussion.  
For simplicity, we consider the incompressible MHD equations with constant mass density $\rho_0$,
\begin{equation}
\partial_t {\u} + {\u}\cdot { {\nabla}} {\u} = - {{ \nabla}} \cpress + \j \times \b
    + \nu \nabla^2 {\u} ,
\label{eq_momentum} \end{equation}
\begin{equation}
\partial_t {\b}  = {{\nabla}} \times (\u \times \b)
    + \eta \nabla^2 {\b}
\label{eq_induction} \end{equation}
\begin{equation}
{{\nabla}} \cdot {\u} =0, \hskip0.2truein {{\nabla}} \cdot {\b} =0
\label{eq_incompressible}
\end{equation}
where ${\u}$ and $\b$ are the velocity and magnetic field 
(in Alfv\'en velocity units,
$\b=\B/\sqrt{\mu_0 \rho_0}$ with $\B$ the induction and $\mu_0$ the permeability); 
$\j = {{ \nabla}} \times \b$ is the current density; 
$\cpress$ is the pressure divided by the mass density; 
and now the normalized viscosity $\nu$ is basically the inverse of $\Rn$, 
and resistivity $\eta$ is basically the inverse of $S$. 

In a dimensionless form, in which all physical quantities are measured by their typical values, 
the MHD equations generally have dissipation terms involving higher spatial derivatives of the field quantities 
with strength characterized by the inverse of dimensionless parameters:  
$\Rn = VL/\nu$ is the Reynolds number 
(with $V$ a typical flow speed, $L$ a typical length scale),   
$S = V_AL/\eta$ is the Lundquist number 
(with $V_A$ a typical \al speed).  
We can immediately see that for most astronomical length scales $L$, 
both $\Rn$ and $S$ are very large numbers such that the dissipation terms in the MHD equations 
can be thought usually to be ignored, (i. e., the ideal MHD equations), 
except possibly in regions where there exist steep spatial gradients. 

In ideal MHD, in which $\nu=\eta=0$ in Eqs. \eq{eq_momentum}-\eq{eq_incompressible}, 
several quadratic quantities, namely 
energy, magnetic helicity and cross helicity are conserved exactly in three dimensions.  
Moreover, a magnetic field line is carried by the flow velocity 
so that the topology of the magnetic field configuration cannot be changed in ideal evolution.  
Thus, many important physical processes, such as magnetic field line reconnection, 
local fluid heating, and particle acceleration due to parallel electric fields, are disallowed.  
In nonideal MHD, such processes can occur within boundary layers, 
which are regions of high spatial gradients in the current density and vorticity, 
described by singular perturbation theory.  
The tendency for formation of current and vortex singularities in the ideal equations, 
if and when it occurs, is a phenomenon of great interest 
because the sites of singularity formation are precisely the sites 
where astrophysically significant physical processes such as heating and particle acceleration can occur.

E. N. Parker 
has argued for over three decades that current sheets, 
or tangential discontinuities of the magnetic field, do generally exist in a magnetic equilibrium
\cite{parker1972, parker1979, parker1994}.  
If Parker is correct, then the formation of current sheets, 
and the subsequent fluid heating and particle acceleration from the release of magnetic energy 
due to the rapid dissipation of these sheets can have very significant astronomical consequences.  
One observable consequence is the production of X-rays.  
The solar corona was the earliest astronomical object that was observed to be emitting X-rays. 
Based on these observations, the temperature of the solar corona is estimated to be of the order of 
a million degrees with peak radiation at wavelengths about 30 A (i. e., in the soft X-ray range). 
While there may be many different heating mechanisms involved in heating the corona, 
it is argued that magnetic fields must play an important role among these processes.
See \cite{nb2008} for our recent discussion on solar coronal heating theory and more references, 
e.g., \cite{Klimchuk2006}.

In Parker's model, a solar coronal loop is treated as a straight ideal plasma column, 
bounded by two perfectly conducting end-plates representing the photosphere.  
The footpoints of the magnetic field in the photosphere are frozen (Òline-tiedÓ).  
Initially, there is a uniform magnetic field along the $z$  direction.  
To simplify the consideration, we may keep the footpoints of the magnetic field on one of 
the plates ($z = 0$ ) fixed, while
the footpoints on the other plate ($z = L$) are subjected to slow, 
random motions that deform the initially uniform magnetic field.  
The footpoint motions are assumed to take place on a time scale much longer 
than the characteristic time for \al wave propagation between $z = 0$  and $z = L$, 
so that the plasma can be assumed to be in static equilibrium nearly everywhere, 
if such equilibrium exists, during such random evolution.  
For a given equilibrium, a footpoint mapping can be defined by following field lines from one plate to the other.  
Since the plasma is assumed to obey the ideal MHD equations, 
the magnetic field lines are frozen in the plasma and cannot be broken during the twisting process.  
Therefore, the footpoint mapping must be continuous for smooth footpoint motion.  
\cite{parker1972} claimed that if a sequence of random footpoint motions renders the mapping sufficiently complicated, 
there will be no smooth equilibrium for the plasma to relax to, 
and tangential discontinuities of the magnetic field (or current sheets) must develop.

Parker's claim has stimulated considerable debate 
\cite{vb1985, antiochos1987, zl1987, ls1994, cls1997} 
that continues to this day.
For review and extensive references, see \cite{low1990, browning1991, parker1994}.  
The first significant objections to Parker's claim of non-equilibrium was raised by van Ballegooijen (1985), 
who argued that smooth equilibria must always exist as long as the footpoint motion is smooth (or continuous).

van Ballegooijen (1985) developed his argument on a reduced form of the MHD equations 
(referred to hereafter as the RMHD equations) originally derived by Strauss (1976) for a low $\beta$ plasma.  
These equations, which also provide the basis for many later developments in this work, are:
\begin{equation}
\partial_t {\Omega} + [\phi,\Omega]  = \partial_z{J} + [A,J]
    + \nu \nabla_{\perp}^2 {\Omega} ,
\label{eq_momentum-RMHD} \end{equation}
\begin{equation}
\partial_t {A} + [\phi,A]  = \partial_z{\phi} + \eta \nabla_{\perp}^2 {A} ,
\label{eq_induction-RMHD} \end{equation}
where $A$ is the flux function so that the magnetic field is expressed as $\b = \lavec{\hat{z}} +  {\nabla}_{\perp}A \times \lavec{\hat{z}}$,
$\phi$ is the stream function so that the velocity field is expressed as $\u =  {\nabla}_{\perp}\phi \times \lavec{\hat{z}}$, 
$\Omega= -\nabla^2_{\perp}\phi$ is the  $z$-component of the vorticity, 
$J = -\nabla^2_{\perp}A$ is the  $z$-component of the current density,  
and $[\phi,A]  = \partial_y\phi\partial_xA - \partial_yA\partial_x\phi$. 
An ideal magnetostatic equilibrium solution of Eqs. (\ref{eq_momentum-RMHD}) and (\ref{eq_induction-RMHD}) 
is obtained by setting all explicitly time-dependent terms, as well as $\phi$ and $\eta$  to zero.  
We then obtain
\begin{equation}
   \partial_z{J} + [A,J] = 0 \, ,
\label{equil-con} \end{equation}
which can also be written as $\b \cdot {\nabla} J = 0$.  
This implies that the current density $J$  must be constant along a given magnetic field-line in an ideal static equilibrium.

Based on this set of equations, 
Longcope and Strauss (1994) argued that even when the magnetic equilibrium is unstable, 
e.g. due to \ci, it will only relax to a second equilibrium (assuming that it exists) 
with very thin current layers with thickness less than about $10^{-7}$ of the large scale.  

However, another point of view was raised by Ng and Bhattacharjee (1998), 
who argued based on a mathematical theorem on the RMHD system that 
for a given fixed footpoint mapping between $z = 0$ and $z=L$, 
there exists only one smooth equilibrium.  
This means that an unstable equilibrium will relax ideally to a final state with current sheets (tangential discontinuities).  
This scenario has very different implications than those predicted by the Longcope and Strauss 1994, 
since energy dissipation and other energetic effects can be much stronger for the case with current sheets, 
than the case with smooth but thin current layers.

Therefore, it is very important to determine which of these two scenarios should actually occur.  
However, due to the fact that the current layers predicted by \cite{ls1994} are very thin, 
it is beyond our computational ability if the full simulation volume is to resolve to the same small scale.  
To provide a resolution to this problem with current computer architectures, 
one may need to apply AMR techniques that put more grid points in the regions where distinct structures appear.  
At the same time, to ensure that any such numerical study is actually representative of the true
dynamic solution, one needs to make sure that the numerical scheme used can maintain a
reasonably high accuracy. Hence, accuracy becomes a particularly important factor in the choice of
the numerical method for the Parker problem of \ci.

Because finite difference--based schemes are usually of low order truncation, accuracy
typically decreases as higher order derivatives are taken. Also because of low
order, these methods can be diffusive (and dispersive).  For the RMHD equations
 (\ref{eq_momentum-RMHD}) and (\ref{eq_induction-RMHD}), one needs to use the spatial 
derivatives of $J$, or third order spatial derivatives of $A$. A couple of questions 
arise in simulations based on such schemes: Will the lower accuracy in calculating these 
higher spatial derivatives change drastically the dynamical properties of the 
problem? Will the numerical diffusion preclude the formation of a current sheet, and lead 
instead to a smooth residual current layer?  For example, in the Parker problem,  one 
will need a simulation that is accurate enough so that we can show confidently whether there 
is a formation of a true current sheet as predicted by Ng \& Bhattacharjee (1998), or if the current 
layers actually tend to fixed (but small) thickness (as small as $10^{-7}$ of 
the large scale) as predicted by Longcope \& Strauss (1994).  In order to increase the reliability of 
the simulations, it is thus of great interest to look for schemes that can maintain higher 
accuracy, and can make use of AMR techniques to resolve 
sharp features at the same time.  Spectral element methods have the potential to fulfill 
such requirements. This is what motivates us to perform the present comparative study.

\section{Simulation set up and diagnostics}
\label{sec_props}

While the main problem of interest is the three dimensional (3D) \ci \ problem 
represented by Eqs. (\ref{eq_momentum-RMHD} )-(\ref{equil-con}), 
for simplicity we will instead simulate the two dimensional (2D) version of the problem 
for the purpose of this comparative study.  
Examining this problem in 2D should not affect greatly differences in accuracy among different codes.  
The 2D \ci\ problem can be viewed as the limit of the 3D problem with $L \rightarrow \infty$.  
In this limit, it has been shown that current singularities must form for the ideal equations ($\eta = 0$) \cite{ls1993}.  
This is beneficial for our present study since we know that there must be specific structures, and we know where they will appear. 
We can then compare how well these sharp structures are resolved in different schemes.  
Therefore, we will simulate the set of equations without the $z$-dependence, or
\begin{equation}
\partial_t {\Omega} + [\phi,\Omega]  = [A,J]
    + \nu \nabla_{\perp}^2 {\Omega} ,
\label{eq_momentum-2D} \end{equation}
\begin{equation}
\partial_t {A} + [\phi,A]  = \eta \nabla_{\perp}^2 {A} ,
\label{eq_induction-2D} \end{equation}
where the definitions for the variables are the same as in Eqs. (\ref{eq_momentum-RMHD} )-(\ref{equil-con}). 

Periodic boundary conditions are used in both the $x-$ and $y-$ directions.  
To be specific, the simulation domain is set to be $0 \leq x \leq 1$ and $0 \leq y \leq 1$.
The initial equilibrium is chosen to be
\begin{equation}
A(x,y,t=0) = A_0 \sin(2\pi x) \sin(2\pi y) \, ,
\label{init-A} \end{equation}
with a small initial flow, which in terms of the stream function is
\begin{equation}
\phi(x,y,t=0) = \phi_0[ \cos(2\pi x) - \cos(2\pi y)] \ \, ,
\label{init-p} \end{equation}
where $A_0=0.4$ and $\phi_0 = 0.002$. 
$\phi_0$ is chosen small enough so that there is a clear linear phase in the growth of the \ci. 
Because SE evolves a different form for the equations (\Sec{sec_sem}),
these initial conditions are converted into conditions 
on $\b$ by using $\b={{\nabla}} \times A \,\lavec{\hat{z}}$, and on $\u$ by using the relation 
$\u={{\nabla}} \times \phi \,\lavec{\hat{z}}$.

\begin{figure}
\plotone{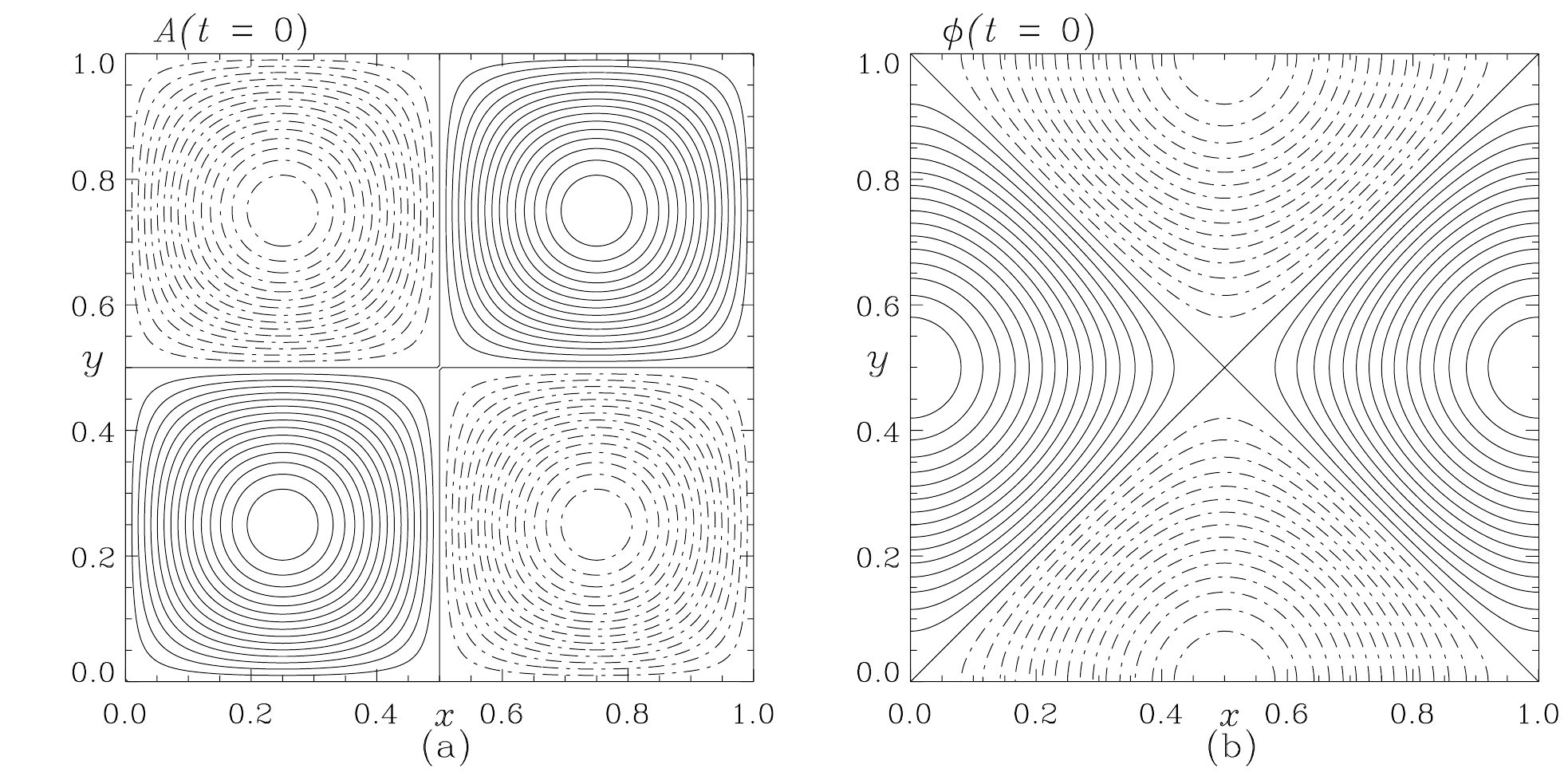}
\caption{Contour plots of (a) the initial flux function $A$, and (b) the initial stream function $\phi$.  
Positive (including zero) contours are solid and negative contours are broken. 
Contours levels are from $-0.4$ to 0.4 with an increment of 0.025 in (a), 
and $-0.004$ to 0.004 with an increment of 0.00025 in (b).
\label{ap0-256}}
\end{figure} 

Fig.~\ref{ap0-256} shows the contour plots of $A$ and $\phi$ at $t=0$ as given by Eqs. (\ref{init-A}) and (\ref{init-p}).
The velocity deduced from this stream function, 
has the initial tendency to push the two islands at the lower right and upper left toward each other and
towards the center.
Eventually, these two islands will merge with each other; hence,  the nomenclature {\it island coalescence}.

The forms in (\ref{init-A}) and (\ref{init-p}) should preserve additional symmetries 
with respect to the lines $x = y$  and $x = 1-y$:  
$A(x,y) = A(-x,-y) = A(y,x)$, $\phi(x,y) = \phi(-x,-y) = -\phi(y,x)$ .  
Specifically,  $\phi = 0$ on the lines  $x = y$  and $x = 1-y$.  
These symmetries are not incorporated into the simulation schemes; however,
they can provide information about how well a numerical scheme preserves them.

The grids are generated separately for each $\Rn$ we consider, and shown in the following
sections. 
The equivalent resolution $\eqr$ for the FD and SE simulations is defined as that which would be achieved 
if the most finely resolved subdomain covered the entire domain. 
 If $E_0$ is the linear number of elements then for SE, equivalent linear resolution,
$\eqr$,  is computed from the initial
$E_0 \times E_0$ grid, the number of refinement levels, $\ell$, and the expansion polynomial
degree, $N$, such that $\eqr = [2^\ell N\, E_0]$. 
For all SE runs, $E_0 = 8$, $N=8$, and $\ell$ varies with $\Rn$.
For the FD method, in order to make more direct comparisons, 
we use the same grids as used in SE, with a $8 \times 8$ uniform grid within each element
when 8$^{\rm th}$ order polynomials are used in SE. 
For the PS method, a uniform grid of $\eqr \times \eqr$ collocation points is used.
Table \ref{tbl_params} contains a list of the viscosity (resistivity) used 
and the corresponding $\eqr$.
\begin{table}
\caption{Parameters used in the simulations described in the following sections; 
         $\eqr$ is the linear grid resolution, and $\nu$ ($\eta$) is the 
viscosity (resistivity).  
         }
\label{tbl_params}
\begin{center}
\begin{tabular}{|c|c|}
\hline
               $\eqr$ & $\nu=\eta$  \\
\hline
               128   & $2\times 10^{-3}$ \\
               256   & $1\times 10^{-3}$ \\
               512   & $3\times 10^{-4}$ \\
\hline
\end{tabular}
\end{center}
\end{table}

In the absence of external forcing, viscosity and magnetic resistivity, 
the 2D MHD equations (\ref{eq_momentum-2D}) and (\ref{eq_induction-2D}) 
have three ideal invariants: 
the energy
\begin{equation}
E = \frac{1}{2} \langle u^2 + b^2 \rangle  = E_K + E_M \, ,
\label{eq_einvar}
\end{equation}
composed of the kinetic and magnetic energy, 
the $L_2$ norm of the magnetic potential
\begin{equation}
M =  \langle A^2 \rangle  \, ,
\label{eq_a2invar}
\end{equation}
and the cross helicity
\begin{equation}
H =  \langle \u \cdot \b  \rangle  \, .
\label{eq_hcinvar}
\end{equation}
For all functions $\phi$ in these definitions, we have
$\langle \phi \rangle  \equiv \int{\phi  \, d{\bf x}^2}$. 

In 2D assuming bi-periodicity, it is easy to show using Eqs.~(\ref{eq_momentum-2D}) --
(\ref{eq_induction-2D}) that
\be
\frac{dE}{dt} = -\nu \langle \Omega^2 \rangle - \eta\langle J^2 \rangle \, ,  
\label{eq_dedt}
\ee
\be
\frac{dM}{dt} = -\eta \langle b^2 \rangle /2\, ,  
\label{eq_da2dt}
\ee
and
\be
\frac{dH}{dt} = -(\eta+\nu) \langle J \Omega  \rangle \, .  
\label{eq_dhdt}
\ee

Equations (\ref{eq_dedt})-(\ref{eq_dhdt}) must hold for {\it any} 2D periodic MHD flow, and they serve
as critical diagnostics for any numerical treatment of the incompressible MHD equations.

In addition to the conservation laws, 
there are several other quantities of interest in 
diagnosing numerical solutions to \ci\, as defined later when we present our numerical results.

\section{Numerical methods}
\label{sec_methods}

We provide in this section the numerical methods we use for carrying out the simulations
on the \ci\ problem. 
These methods have been described in detail in other sources, 
but we provide enough explanation so that the simulation setup, 
results, and discussion will be comprehensible, and the paper reasonably
self--contained.

\subsection{Spectral element method}
\label{sec_sem}

This method evolves Eqs.~(\ref{eq_momentum})-(\ref{eq_incompressible}) in time, as part
of the {\it Geophysics/Astrophysics Spectral Element Adaptive Refinement (GASpAR)} code, and 
has been described in detail in \cite{rosenberg2007}.
Here we present aspects of that description that relate to the results discussed in this
work. 
Equations \eq{eq_momentum}-\eq{eq_incompressible} are solved in Els\"asser form \cite{elsasser1950}:
\be
\partial_t \Z^{\pm} + \Z^{\mp}\cdot {{\nabla}} \Z^{\pm} + {{\nabla}} \cpress -\nu^{\pm}\nabla^2 \Z^{\pm} - \nu^{\mp} \nabla^2 \Z^{\mp} = 0
\label{eq_zmomentum}
\ee
\be
 {{\nabla}} \cdot  \Z^{\pm} = 0\,,
\label{eq_zconstraint}
\ee
where $ \Z^{\pm} = \u \pm \b $ and $ \nu^\pm = \frac{1}{2}(\nu \pm \eta)$.
Thus, we solve for $\u$ and $\b$ in terms of $\Z^{\pm}$. All components of $\Z^{\pm}$
and those of the velocity and magnetic field reside on Gauss--Lobatto--Legendre (\gll) nodes,
while pressure resides on Gauss--Legendre (\gl) nodes. These choices follow from the 
finite dimensional subspaces to which these quantities are restricted:
 $\Z^{\pm}$ (and $\u$ and $\b$) are expanded in terms of $N^{\rm th}$ order \gll\ polynomials, while
$\cpress$ is expanded in terms of $N-2^{\rm th}$ order \gl\ polynomials. 
Substituting these expansions into the $d$-dimensional variational form of equations
(\ref{eq_zmomentum})-(\ref{eq_zconstraint}) on a domain $\pdomain$, and using appropriate 
quadrature rules, we arrive at a set of semi--discrete equations written in terms of
spectral element operators:
\begin{eqnarray}
\Mmat \frac{d\zpmnum_j}{dt} & = & -\Mmat\amat^{\mp}\zpmnum_j + \DTmat_j \ppmnum 
\nonumber \\
& & -\nu_{\pm}\Lmat\zpmnum_j -\nu_{\mp}\Lmat\zmpnum_j 
\label{eq_zpsemidiscrete}\\
\Dmat^j \zpmnum_j & = & 0,
\label{eq_divsemidiscrete}
\end{eqnarray}
for the $j^{\rm th}$ component. 
The variables ${\znum}^\pm$ represent values of the $\Z^\pm$ collocated at the GL node points, and
$\pnum^\pm$ are values of the pressures at the G node points. Note that, even though the continuous
equations contain only a single pressure, Eq.~(\ref{eq_zpsemidiscrete}) contains one for {\it each} 
Els\"asser vector because the constraints (\ref{eq_divsemidiscrete}) are enforced separately on them.
The operators $\Mmat$, $\Lmat$, and $\amat$, are the well--known
mass matrix, weak Laplacian and advection operators, respectively (\eg, \cite{rosenberg2006}, and references
therein), and $\Dmat_j$ represent the Stokes operators, in which the \gll\ basis function and its derivative
operator are interpolated to the \gl\ node points, and multiplied by the \gl\ quadrature weights.
All $d$--dimensional operators are computed as tensor products of their component 1D operators. Note
that because different expansion bases are used for the vector quantities as are used for
pressure, a staggered grid is implied. Hence, this method is referred to as the $\polynomialsset{N}-\polynomialsset{N-2}$
formulation. It was chosen to prevent spurious pressure modes \cite{maday1992,fischer1997}.
Note also the effect of the $\Dmat_j$, which act on so--called {\it v-grid} (vector) quantities: they take a 
derivative that itself resides on the \gl\ nodes; hence, the discrete divergence, Eq. \eq{eq_divsemidiscrete}
resides on the p-grid. The effect of the 
transposed Stokes operator, $\DTmat_j$, on the other hand, is to compute a derivative--residing on the v--grid--of a {\it p-grid} quantity.

The code has an adaptive mesh capability that we utilize minimally in this work. The connectivity and
adaptivity algorithms were described in \cite{rosenberg2006}. No dynamic adaptivity is used here
(see \Sec{sec_props}). Instead, the nonconforming grid is constructed initially by turning off
the refinement criteria, and selecting the elements we want to refine explicitly. The nonconforming
grid is then used in a static configuration throughout the simulation. {\it Nonconforming} in this 
context means that there are at most two child elements adjacent to a coarser neighbor; 
an element is refined by dividing it isotropically into $2\times2$ child elements, each of which contains
the same polynomial order as the parent.

\subsubsection{Time stepping with a constraint}

In our simulations we must resolve all temporal (and spatial) scales, so we use
an explicit Runge--Kuttta (RK) method for integrating Eqs. \eq{eq_zpsemidiscrete}-\eq{eq_divsemidiscrete} in time.
The specific RK algorithm is that presented in \cite[p. 109]{canuto1988}, known to 
be valid to second order in  $\Delta t$ \cite{brachet2007}, which we use for all computations. At each stage,
we can write (recall eq. \eq{eq_zpsemidiscrete}): 
\begin{eqnarray}
\znum_{j}^\pm & = & \znum_{j}^{\pm,n}  - \frac{1}{k} \Delta t \; \Mmat^{-1}(\Mmat\amat^{\mp} \znum_j^\pm  -  \DTmat_j \pnum^\pm  \nonumber \\
& & +\nu_{\pm}\Lmat\znum_j^\pm +\nu_{\mp}\Lmat\znum_j^\mp ) .
\label{eq_rkstage}
\end{eqnarray}
We require that each RK stage obey eq. \eq{eq_divsemidiscrete} in its discrete form, so multiplying
eq. \eq{eq_rkstage} by $\Dmat_j$, summing, and setting the term $\Dmat^j \znum_j^\pm = 0$
leads to the
following pseudo-Poisson equation for the pressures, $\pnum^\pm$:
\be
\Dmat^j \Mmat^{-1} \DTmat_j \pnum^\pm = \Dmat^j \gnum_j^\pm,
\label{eq_pseudopoisson}
\ee
where the quantity
$$
\gnum_j^\pm =  \frac{1}{k} \Delta t \; \Mmat^{-1}(\Mmat \amat^{\mp}\znum_j^\pm  + \nu_{\pm}\Lmat\znum_j^\pm +\nu_{\mp}\Lmat\znum_j^\mp ) - \znum_j^{\pm,n}
$$
is the remaining inhomogeneous contribution. Rosenberg et al. (2007) describe how $\Mmat^{-1}$ is computed.
Equation \eq{eq_pseudopoisson} is solved using
a preconditioned conjugate gradient method (PCG); for all computations reported here, 
we use a block Jacobi preconditioner computed using a fast diagonalization method.

Thus, at each time-step, two RK stages are computed, and each stage solves eq. \eq{eq_pseudopoisson} twice,
once for $\zpnum$, and once for $\zmnum$ leading to $4$ pseudo-Poisson solves at each time step.

\subsubsection{Communication}

While equations \eq{eq_zpsemidiscrete}-\eq{eq_divsemidiscrete}
are correct for a single element, strictly speaking, they are not complete when 
multiple subdomains (elements) are used. In this case, we must ensure that all quantities in 
the subspace represented by the \gll\ grid remain
continuous across element interfaces. The specific way in which this is carried out in the code
is described in \cite{rosenberg2006}, and entails exchanging interface data.
Using the Boolean scatter matrix, $\conlab{\Qmat}$, the interpolation
matrix from global to local degrees of freedom, $\Jmat$,  and the masking matrix (that
enforces homogeneous boundary conditions), $\mask$, that were presented there (see also Fischer et al. 2002 ),
it is found that the local Stokes operators in equations \eq{eq_zpsemidiscrete}-\eq{eq_divsemidiscrete} must be 
replaced with 
$$
\Dmat_j \to \Dmat_{L, j} \Jmat\conlab{\Qmat}\mask, 
$$
where $\Dmat_{L, j}=\diag_k(\Dmat_{j}^{k})$, and the $\Dmat_{j}^{k}$ are the
matrices from above, and $k$ indexes the elements in the domain. It is sufficient for our
purposes to refer to the local form of the Stokes operators, and to simply observe that communication
occurs when they are applied. 

\subsection{Finite difference method}

The {\em Magnetic Reconnection Code (MRC)} is a suite of
codes \cite{astronum2006,b2005} which integrate various reduced and extended
fluid models of plasma flows. 
In this paper, the 2D AMR version of
the code integrating the equations of RMHD has been used.

The MRC employs a hierarchical quadtree based approach in
block-structured adaptive mesh refinement. 
At each refinement time,
every (square or rectangular) box is checked as to whether the data in
that box requires additional resolution. 
The refinement criterion needs to be selected appropriately for the given problem, 
from simple evaluation of maxima or gradients to a Richardson extrapolation based approach.

If the local resolution in a box is considered insufficient, 
it is subdivided into $2 \times 2$ smaller boxes, 
which have physically twice the resolution but the number of grid points in each box remains
the same it was on the coarse parent box.

As opposed to the alternate approach of allowing for arbitrary rectangular patches of refinement, 
this method results in simpler data management, 
easier optimization (each box is always the same size of $n \times m$ grid points, 
important for cache considerations, etc) 
and more efficient load balancing, 
even though it introduces some inefficiency in that some additional areas are refined 
where the higher resolution is not required.

For load-balancing, we use a space-filling Hilbert Peano curve which
connects all the boxes at various levels of refinement. 
The boxes along this one-dimensional curve are then evenly distributed to the
available processors. 
Since each patch has the same number of grid points, 
the computational load is evenly spread, 
and as the Hilbert-Peano curve has the property that in a certain averaged sense,
patches which are close in the two-dimensional domain are also close
on the one-dimensional curve, 
spatially close regions are clustered onto the same processor; 
that is, it maintains data locality.
Since communication is only necessary between spatial neighbors to exchange boundary data, 
most of the needed data is available on the local processor, 
expensive MPI communication is only required for boundary data transfer across the boundaries
between clusters on different processors, which is effectively minimized.

An additional difficulty in using AMR to integrate the reduced models -- not encountered
 in  the  purely hyperbolic systems -- is the need to solve
elliptic equations, e.g. solving for the stream function $\phi$ from the vorticity $\Omega$. 
To discretize this sub-problem, 
we rewrite the Laplacian $\nabla^2$ as the divergence of a gradient ${\nabla}\cdot{\nabla}$ and
apply a conventional finite volume method for a conservation law.
Combined with correcting the fluxes at fine-coarse boundaries, 
this provides an exactly numerically conservative discretized expression.
To efficiently solve this elliptic problem we use a variation of the
fast adaptive composite (FAC) method \cite{g2004,m1989}, 
which, due to the multilevel character, provides an iterative solver with very fast
convergence.

Alternatively, the discretized Laplacian on the AMR hierarchy can be
calculated explicitly as a sparse matrix, and PETSc's \cite{petsc} rich
supply of solvers are available to solve Poisson's equation. 
In particular, SuperLU \cite{superlu_dist} in many cases proves to be very fast, 
since the expensive LU decomposition only needs to be done once
and then can be applied for many time-steps until the AMR hierarchy of
boxes changes. 
This is the method used in the simulations for this study.

Our AMR code can also be run in fully implicit mode using
Crank-Nicholson time-stepping, however no preconditioner has been
developed yet, so we are using a simple unpreconditioned Newton-Krylov
solver which does not achieve optimal performance, but is still faster
than a Runge-Kutta explicit scheme.
Most of the runs presented here are using the implicit time
integrator. 
Some runs have also been cross-checked by using an adaptive time-step
Runge-Kutta explicit method, which is included with PETSc's.

Spatial discretization is using a regular second order central differences, 
which is equivalent to a conservative finite volume method, 
with no upwinding employed.

For the purpose of the present study, the dynamic refinement capability 
of the AMR code is turned off so that we can use the same grids that were also used
in SE. This was done because it is difficult to set refinement methods and criteria to be the
same in both the FD and SE codes. Using static refinement provides us with nonconforming
grids over which we can exert complete control of the number of degrees of 
freedom. Comparison of refinement criteria in the SE and FD codes and the effect on the
resulting dynamics is left for future work.

\subsection{Pseudo-spectral method}

The PS code is based on fast Fourier transform (FFT) on a 2D bi-periodic domain. 
It is de-aliased by the standard 2/3 rule.  
The nonlinear term is calculated in the physical space on a uniform grid of collocation points.
A second order predictor-corrector method is used for time integration. 
The code is parallelized using a parallel version of the FFT.  
For this study,  results from SE and FD methods for the $\eqr$
case are compared with those from the PS code with $\eqr \times \eqr$ collocation points.
The results from the PS code are themselves checked with PS runs with higher
resolutions, up to $2048 \times 2048$,  which confirm that the results with the original
resolution are already well resolved.

\section{Numerical results}
\label{sec_results}

\begin{figure}
\plotone{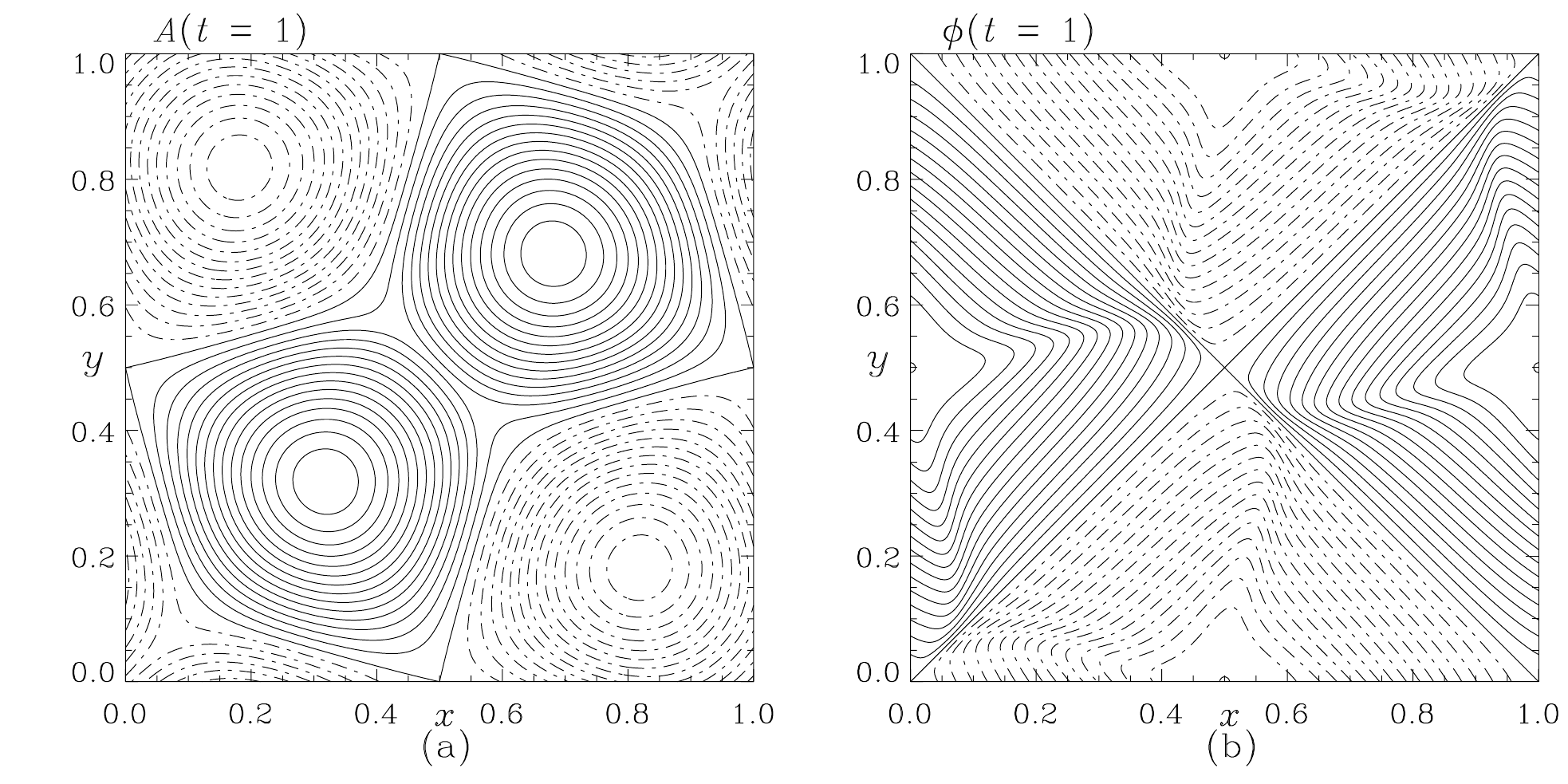}
\caption{Contour plots of (a) $A$, and (b) $\phi$ at $t = 1$ for the $\eqr = 256$ case.  
Positive (including zero) contours are solid and negative contours are broken. 
Contours levels are from -0.4 to 0.4 with an increment of 0.025 in (a), 
and -0.1 to 0.1 with an increment of 0.00625 in (b). Note the sheet formation (a) and the 
fluid acceleration (b).
\label{ap1-256}}
\end{figure} 

In this section, we compare simulation solutions by SE and FD methods, using solutions by PS as references.
Note that each code (SE and FD) has been verified separately on a suite or problems including analytical
solutions, but the purpose of the present work is to explore the fully developed nonlinear
regime of the \ci\ problem, when sharp current and vorticity layers appear.

The main dynamics of \ci\  is due to the attractive force between two islands with the same sign of current. 
The initial small perturbation in the flow velocity breaks the unstable equilibrium.
After a short transient, a linear phase appears when kinetic energy increases exponentially so that
the flow velocity pushes the two islands together further. 
Eventually a sharp current layer appears between the two islands when the dynamics enters the nonlinear phase
at time $t \sim 1$, in units of the \al time.
Magnetic reconnection then proceeds faster, producing a strong outflow velocity, as well
as sharp vorticity layers.
Fig.~\ref{ap1-256} shows contour plots of $A$ and $\phi$ at $t = 1$ for the $\eqr = 256$ case.
We see that the two islands with positive $A$ are pushed towards each other 
with some $A$ contours (magnetic field lines) already reconnected, as compared to Fig.~\ref{ap0-256}.
The stream function $\phi$ shows strong outflows as indicated by the concentration of stream lines.

\begin{figure}
\includegraphics[width=2.8in]{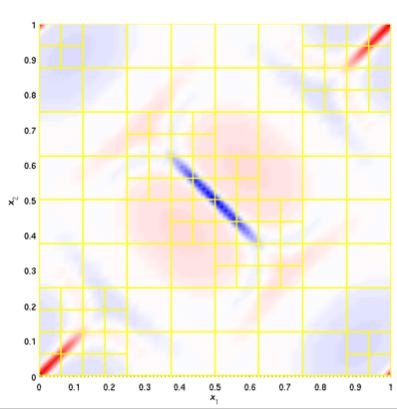}
\caption{The statically refined grid used in the $\eqr = 128$ case. The background is the color contour plot of the current density, $J$, at $t = 1.3$ produced by the SE run, with red representing the positive end and blue representing the negative end of $J$ values.  Within each square, polynomials of the order of $N = 8$ are used for SE, while an $8 \times 8$ uniform grid is used for FD. Only one level of refinement is needed at this (low) Reynolds number.
\label{128-grids}}
\end{figure} 

\begin{figure}
\includegraphics[width=2.8in]{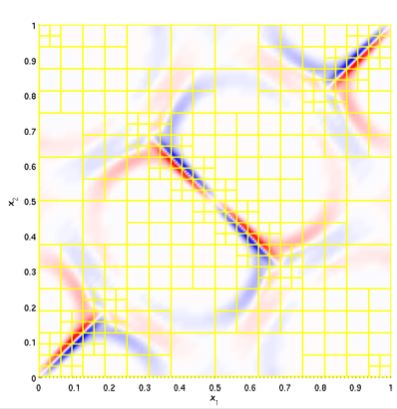}
\caption{The grid used in the $\eqr = 256$ case. The background is the color contour plot of the vorticity $\Omega$ at $t = 1.3$ produced by the SE run, with red representing the positive end and blue representing the negative end of $\Omega$ values. Two levels of refinement are used here.
\label{256-grids}}
\end{figure} 

\begin{figure}
\includegraphics[width=2.8in]{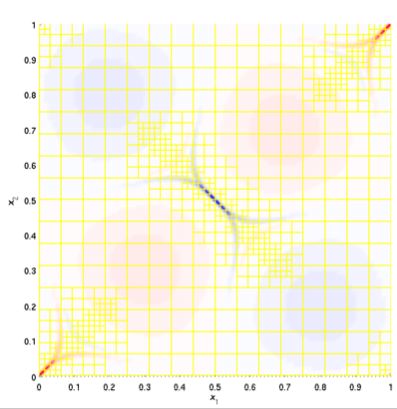}
\caption{The grid used in the $\eqr = 512$ case. The background is the color contour plot of the current density $J$ at $t = 0.93$ produced by the SE run, with red representing the positive end and blue representing the negative end of $J$ values. Now, three levels of refinement are used.
\label{512-grids}}
\end{figure} 

The grids used in our simulations for the three cases are shown in Fig.~\ref{128-grids} to \ref{512-grids}.  
The refinements with each grid are imposed so as to resolve structures produced by the above dynamics, for the different dissipation levels indicated in Table \ref{tbl_params}.  
In Fig.~\ref{128-grids}, we also plot the color contours of the current density $J$ at $t = 1.3$ 
produced by the SE run for the  $\eqr = 128$ case,
with red representing the positive end and blue representing the negative end of $J$ values (as in the following).  
Only one level of refinement is used that covers the region containing stronger $J$.  
Note that within each square, polynomials of order $N = 8$ are used for SE, while an $8 \times 8$ uniform grid is used for FD (the same as in the other two cases).
In Fig.~\ref{256-grids}, the color contours of the vorticity $\Omega$ at $t = 1.3$ produced by the SE run are plotted along with the grid for the $\eqr = 256$ case.
Note that  there are now two levels of refinement.
In Fig.~\ref{512-grids}, the color contours of $J$ at $t = 0.93$ produced by the SE run are plotted along with the grid for the $\eqr = 512$ case, with one further level of refinement.
At the same time, we found that the largest squares have to be refined for this case due to a much smaller dissipation in this case.

\begin{figure}
\plotone{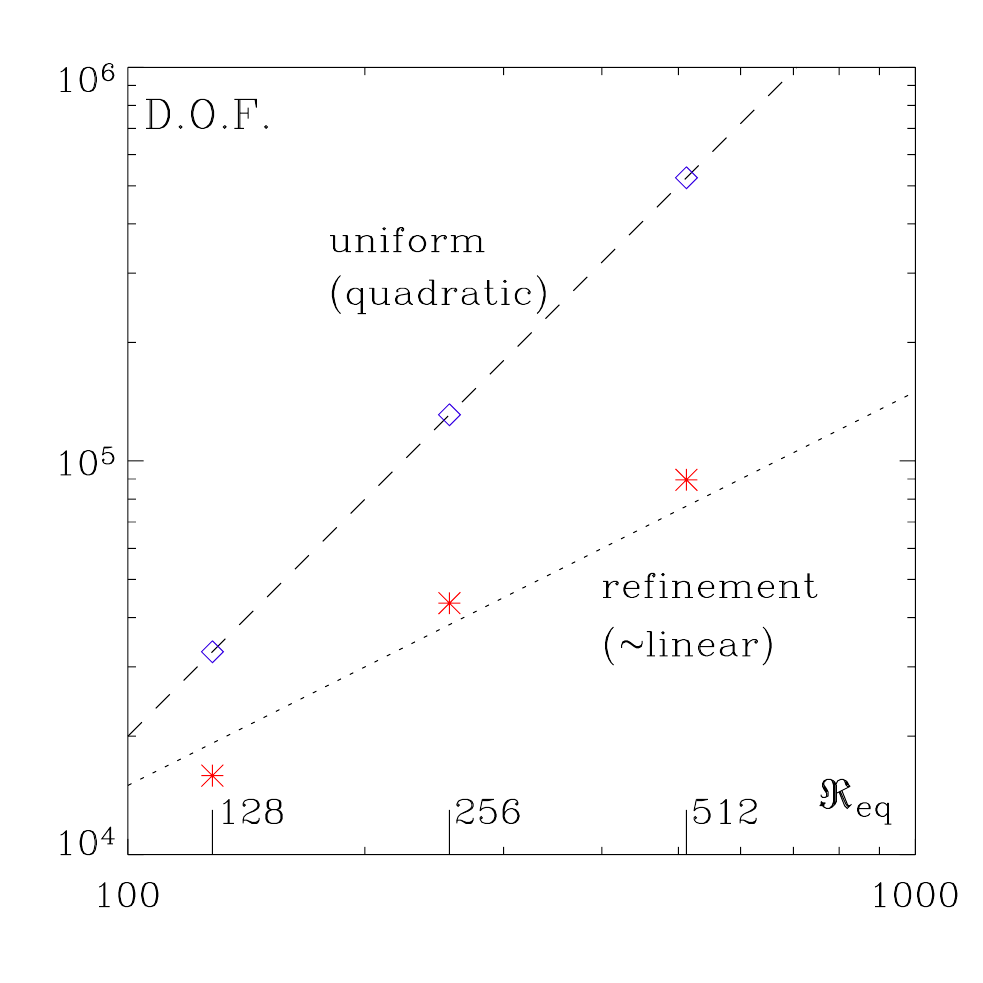}
\caption{DOF of the grids shown in Fig.~\ref{128-grids} to \ref{512-grids} at the different $\eqr$ levels, in red asterisks,
and those of the PS runs using uniform grids, in blue diamonds.  
A dashed line showing a $\eqr^2$ dependence in the PS runs,  
and a dotted line showing a linear dependence of $\eqr$ of the statically--refined runs are also plotted.
\label{dof}}
\end{figure} 

\begin{figure}
\plotone{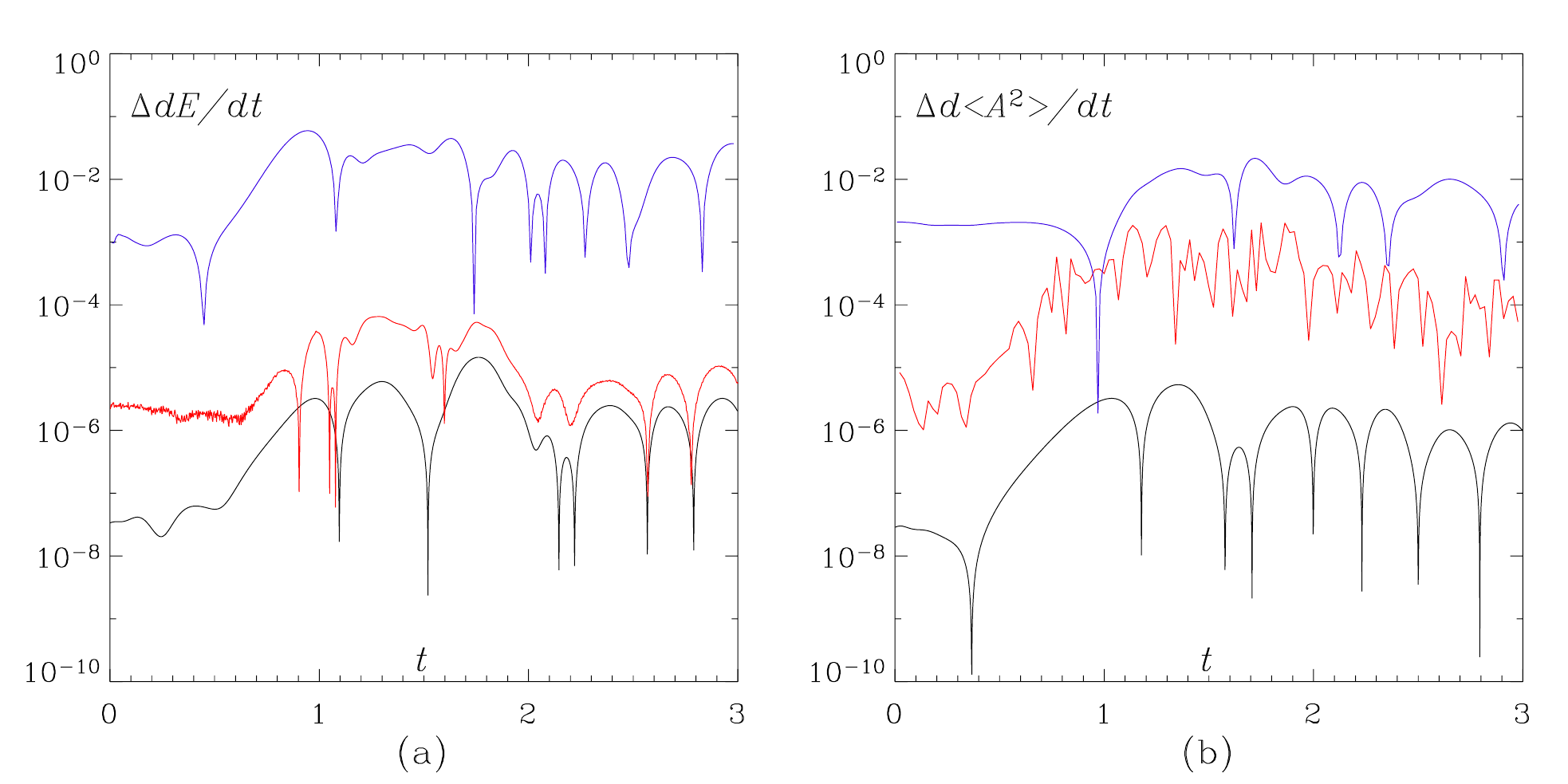}
\caption{Fractional error in energy conservation law in (a) $\Delta \dot{E}$, and for the $\Ltwo$ norm of the 
magnetic potential, $\Delta \dot{M}$ in (b) for the $\eqr = 128$ case
as functions of time.
 Black curves are for the PS run, red curves are for the SE run, and blue curves are for the FD run 
 (same in the other figures below) . 
\label{c128}}
\end{figure} 

\begin{figure}
\plotone{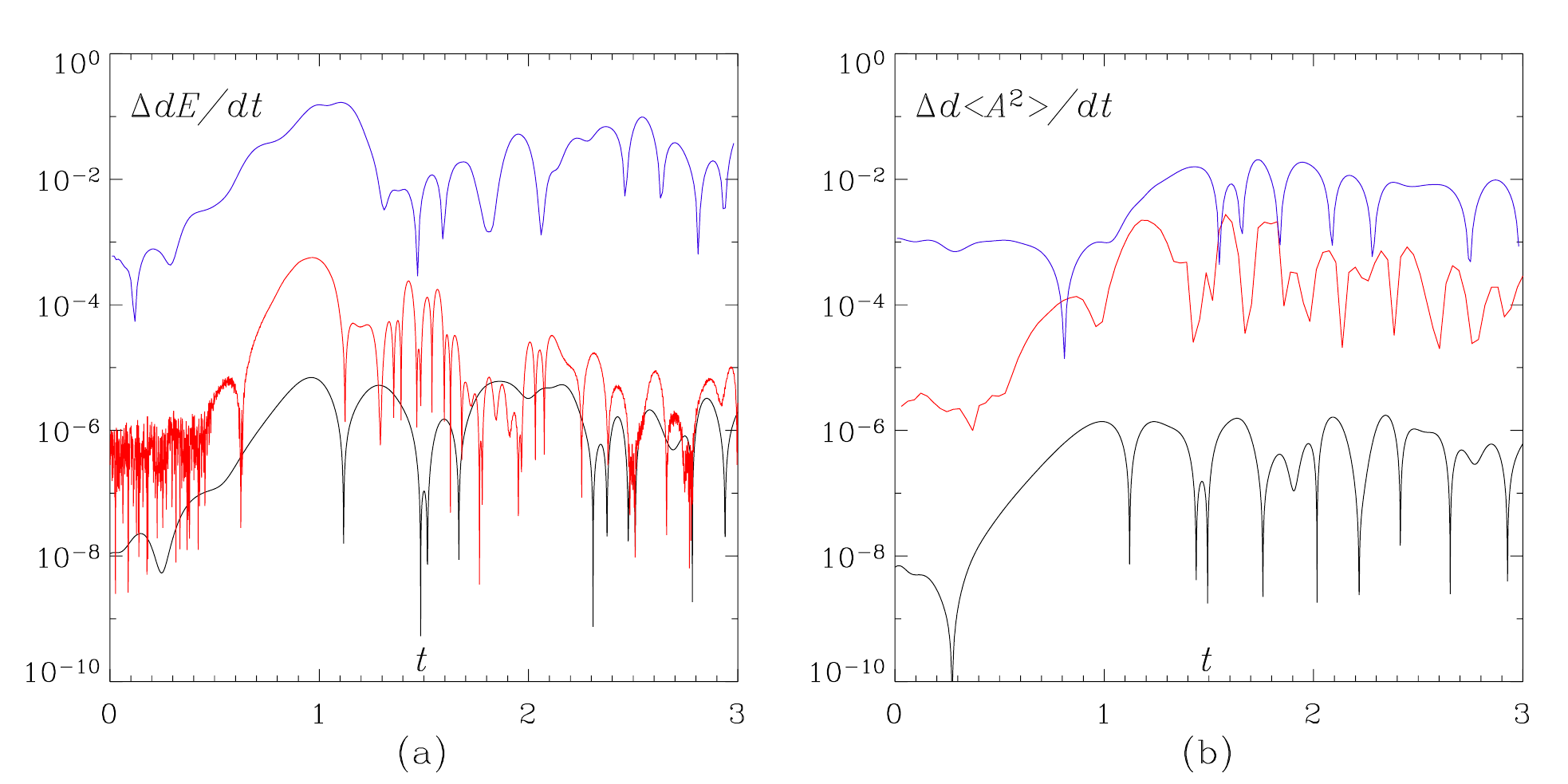}
\caption{Fractional error in conservation laws in (a) $\Delta \dot{E}$, and (b) $\Delta \dot{M}$ for 
the $\eqr = 256$ case as functions of time. See Fig. \ref{c128} for definitions.
\label{c256}}
\end{figure} 

\begin{figure}
\plotone{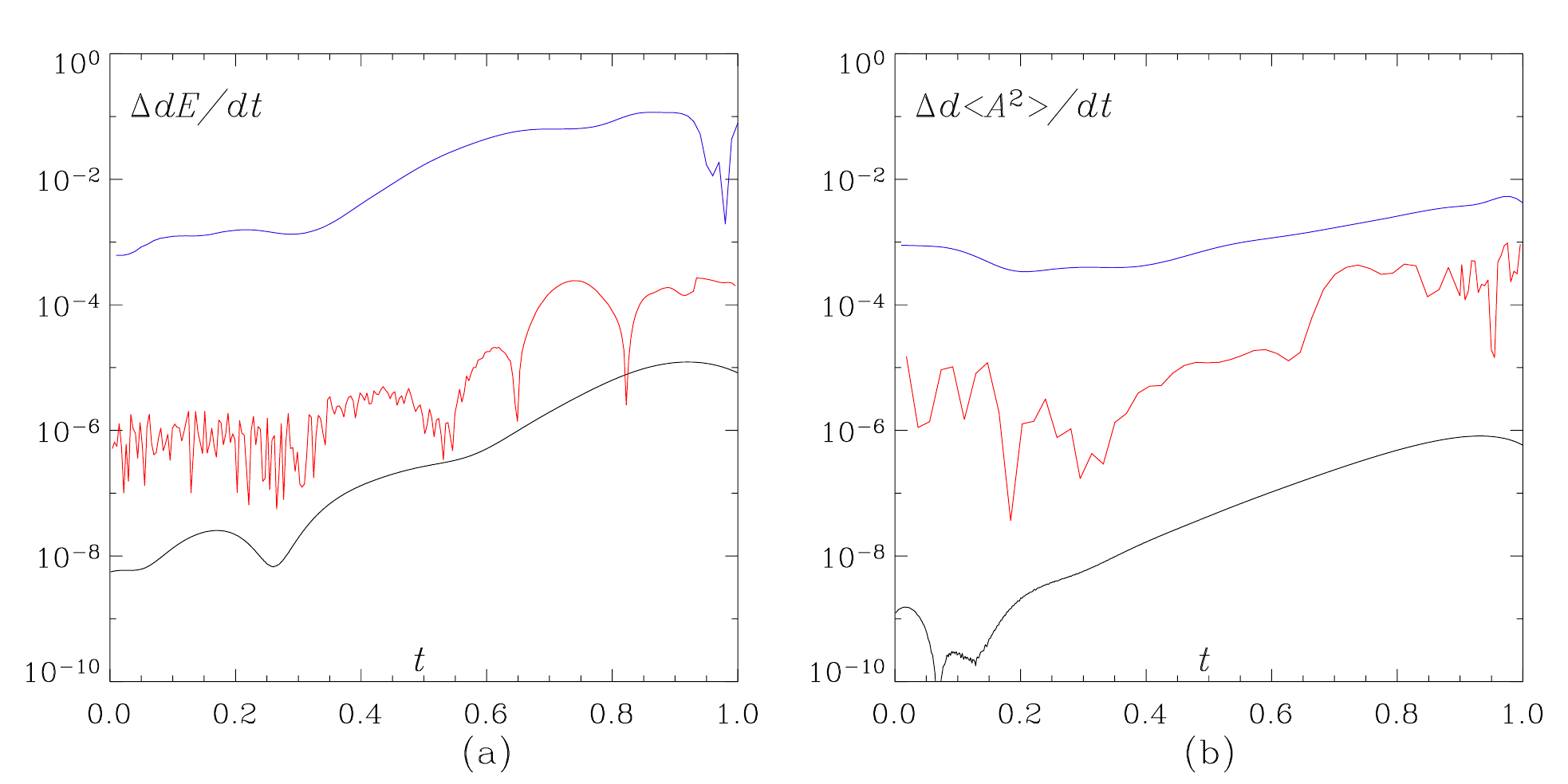}
\caption{Fractional error in conservation laws in (a) $\Delta \dot{E}$, and (b) $\Delta \dot{M}$ for 
the $\eqr = 512$ case as functions of time. See Fig. \ref{c128} for definitions.
\label{c512}}
\end{figure} 

Before presenting our numerical results, let us look at the degrees of freedom (DOF) of these grids at the different $\eqr$ levels,
as shown in Fig.~\ref{dof} in red asterisks; they follow roughly a linear scaling proportional to $\eqr$.
Also plotted are DOF of the PS runs using uniform grids, in blue diamonds, 
which follow a scaling of $\eqr^2$ as they should.
Since as $\eqr$ increases, the difference between the two scalings can be very large,
using adaptive grid refinement has the potential to provide considerable savings in memory
and/or CPU usage, if the linear scaling of the DOF in these adaptive grids continues to
hold for even larger $\eqr$, and thus for large $\Rn$ and $S$. 

\subsection{Accuracy of the conservation laws}
\label{sec_conserve}

We start our comparison by looking at how each code preserves conservation laws as shown in Eqs.~(\ref{eq_dedt}) and (\ref{eq_da2dt}).  
We do not include the conservation law for cross helicity, Eq.~(\ref{eq_dhdt}), in our comparison, 
since $H = 0$ exactly based on our initial conditions in Eqs.~(\ref{init-A}) and (\ref{init-p}), 
and thus $dH/dt = 0$; all three codes preserve  $H = 0$ and $dH/dt = 0$ well during the 
duration of the runs for all three cases.
This is more due to how well the structure of each code preserves symmetries,
rather than due to numerical accuracy. Also, since  both the left hand side (LHS) and right hand side
(RHS) of Eq.~(\ref{eq_dhdt}) are close to zero,
taking the difference between the two to see fractional changes will not yield meaningful results.

In Figs.~\ref{c128} to \ref{c512}, the fractional difference between the LHS and RHS of (a) Eq.~(\ref{eq_dedt}), 
and (b) Eq.~(\ref{eq_da2dt}) are plotted as functions of time, 
for the three values of $\eqr$.
The fractional difference (or error) $\Delta$ is defined as $\mid {\rm LHS} - {\rm RHS} \mid / \mid {\rm RHS} \mid$,
with the time derivative in the LHS calculated by taking finite difference of the output in time, thus 
providing an overestimate of the error. 
In these figures, black curves are for PS runs, red curves are for SE runs and blue curves are for FD runs. 

For the $\eqr = 512$ case, we present results up to $t \sim 1$, since both SE and FD experience larger
error after this time, 
probably due to the fact that we are using a fixed adaptive grid and could not follow the development
of small scales closely enough.
To achieve more accurate results, both codes would have to employ dynamic adaptive refinement.
However, it is difficult to make certain the two codes refine and coarsen in the same way in order 
to make meaningful comparisons; hence, this investigation is left for future work.

For all three levels of $\eqr$, we see that PS results (black) preserve conservation laws the best, as expected.
This is why we may use them as baselines for comparisons.
The error level in $\Delta \dot{M}$ as shown in (b) panels 
(with $\dot{M} \equiv d\langle A^2 \rangle/dt$) is around or slightly over $10^{-6}$,
while the error $\Delta \dot{E}$ as shown in (a) panels is around or slightly below $10^{-5}$,
since energy involves one more spatial derivative than $A$ and thus is less accurate in this
computation.

For SE runs (red), the accuracy turns out to be quite good, given that they are running on
statically refined grids.
The error $\Delta \dot{E}$ is more or less one order of magnitude above that of PS runs for all three cases,
but still is in a low level of about $10^{-4}$.  
For the error $\Delta \dot{M}$, the difference between SE and PS becomes larger (about two orders of magnitude), at a level of about $10^{-3}$.  
The reason $\Delta \dot{M}$ is greater than $\Delta \dot{E}$ for SE runs is due to the fact that
$\b$ and $\u$ are the primary variables in the computations,
and $A$ is a field quantity derived from $\b$, by solving the equation $\nabla^2 A = -J$.
This process introduces error in $A$ and thus the error $\Delta \dot{M}$ is greater than $\Delta \dot{E}$.

For FD runs (blue), the code does capture quantitatively the main dynamical evolution of the \ci\ problem, although with a lower accuracy.
For $\Delta \dot{M}$, the error level is up to above $10^{-2}$, and the error $\Delta \dot{E}$ can go above that, even reaching as high as $10^{-1}$ or more.
The error in $\Delta \dot{M}$ is at a level slightly lower than that of $\Delta \dot{E}$ in this version of the FD code, similar to the trend of PS, since it uses $A$ as primary variable.

From these comparisons of the conservation laws, we see that the accuracy of SE is higher 
as expected for spectral--based methods; it sometimes approaches that of PS runs using uniform grids. 
This confirms that SE can deliver near spectral accuracy, the main advantage of employing
 such a scheme. Using the same adaptive grid, FD runs show lower accuracy.
It is conceivable that the accuracy of the FD scheme can be improved somewhat by algorithm
modifications. However, it is unlikely that it can be improved to the level of the SE runs, 
when constrained to the same grids.

\begin{figure}
\plotone{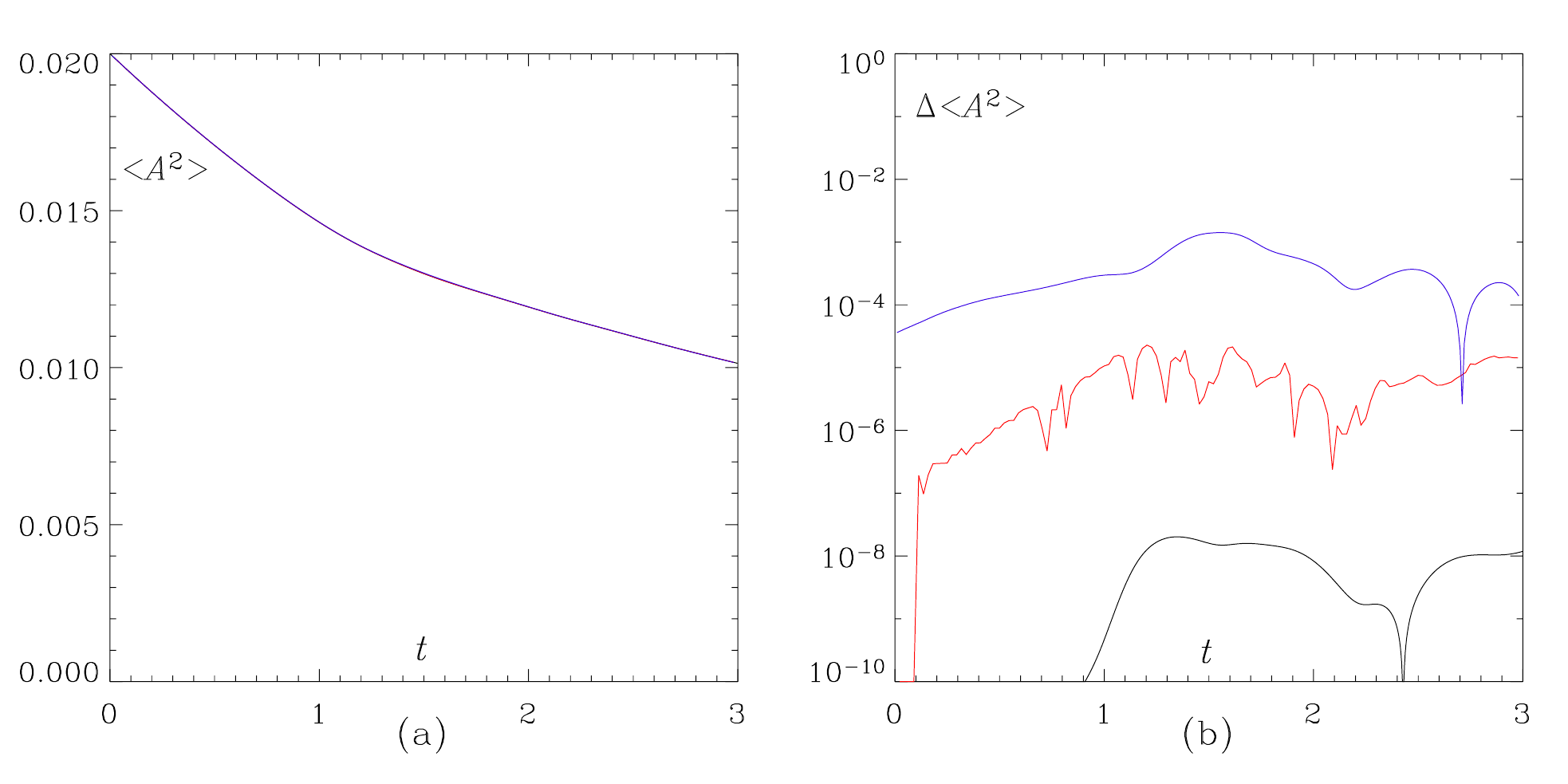}
\caption{(a)  Time series of $\langle A^2 \rangle$ for the PS (black), SE (red), and FD (blue) runs 
plotted in this order for the $\eqr = 128$ case. All three runs overlap.
(b) Fractional error in $\Delta \langle A^2 \rangle$, as compared with a PS run with $2048 \times 2048$ resolution.
\label{a2-128}}
\end{figure} 

\begin{figure}
\plotone{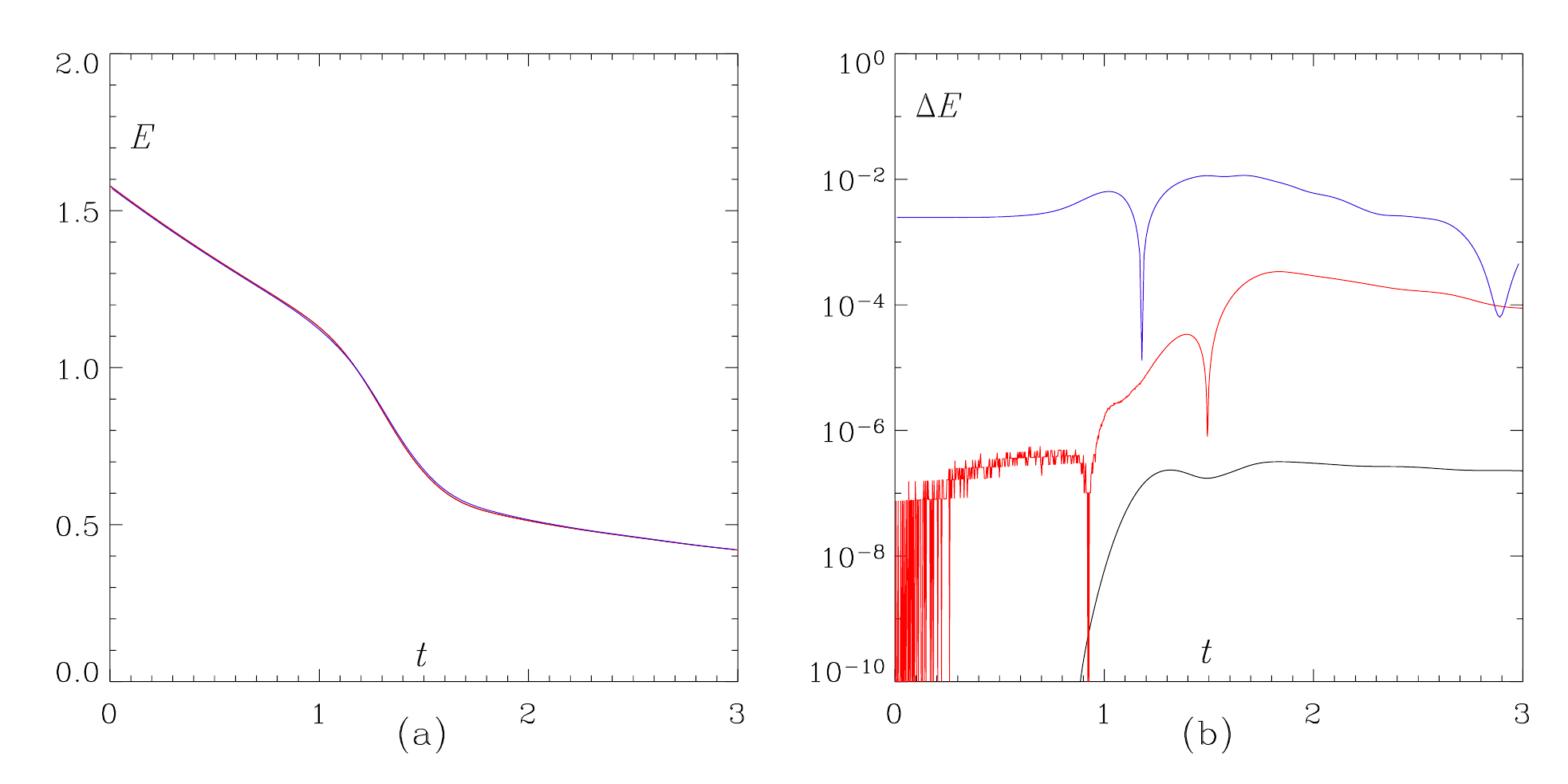}
\caption{(a)  Time series of total energy $E$ for the PS (black), SE (red), and FD (blue) runs 
plotted in this order for the $\eqr = 128$ case. Again, all three runs are nearly coincident.
(b) Fractional error in $\Delta E$, as compared with a PS run with $2048 \times 2048$ resolution.
\label{e-128}}
\end{figure} 

\begin{figure}
\plotone{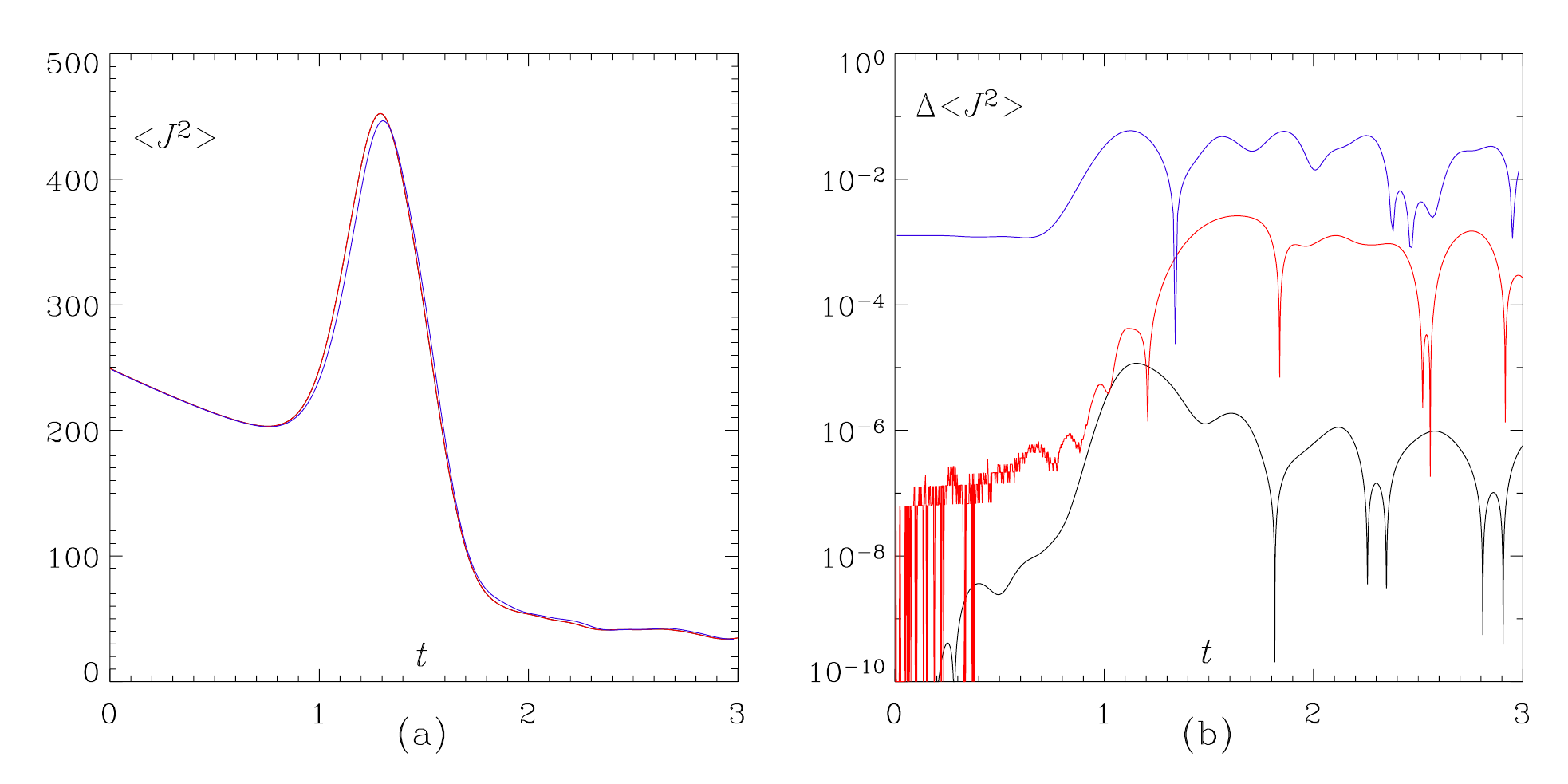}
\caption{(a)  Time series of $\langle J^2 \rangle$ for the PS (black), SE (red), and FD (blue) runs 
plotted in this order for the $\eqr = 128$ case. Differences in the dissipation $\eta \langle J^2 \rangle$
are observed.
(b) Fractional error in $\Delta \langle J^2 \rangle$, as compared with a PS run with $2048 \times 2048$ resolution.
\label{j2-128}}
\end{figure} 

\begin{figure}
\plotone{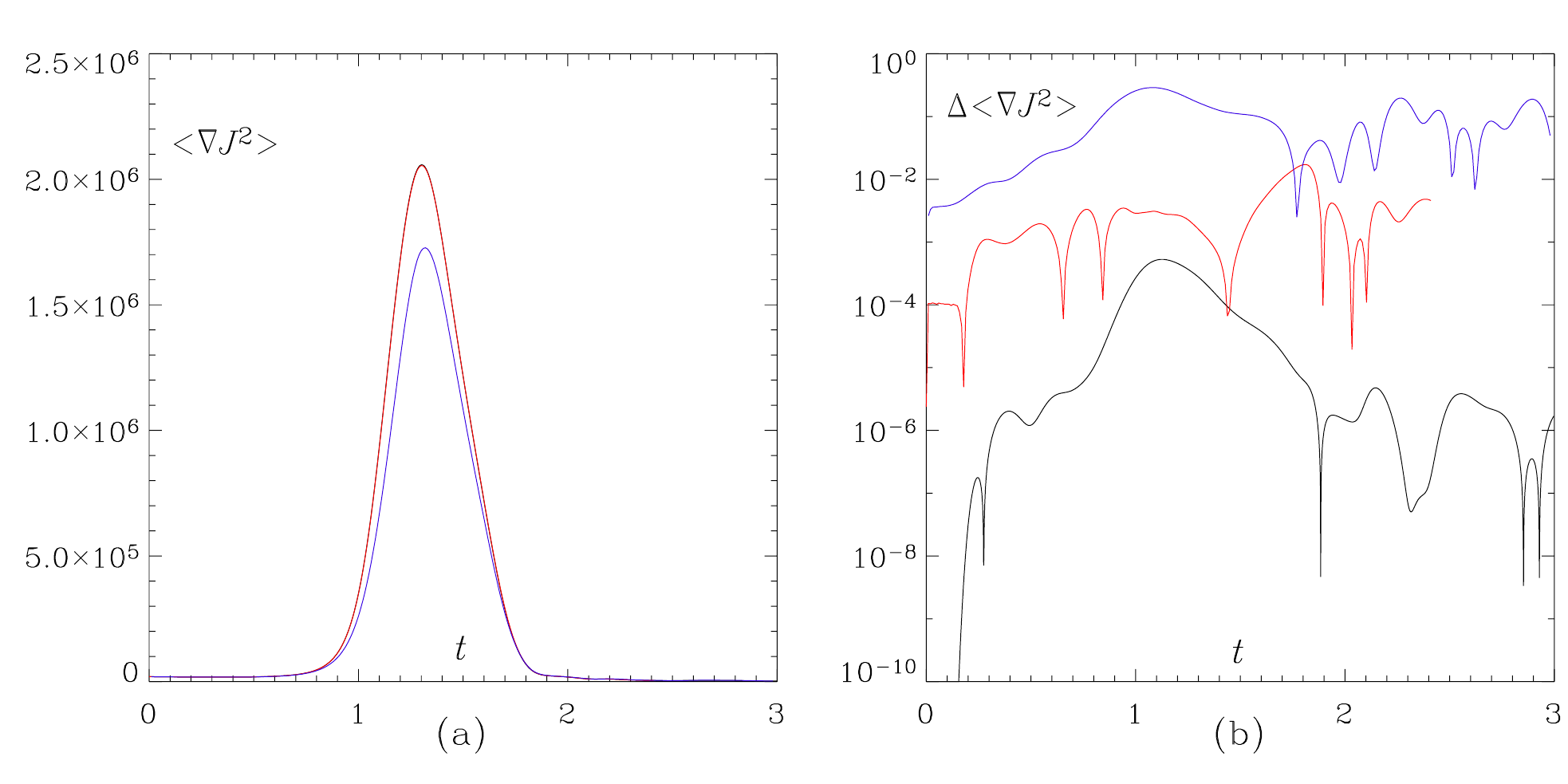}
\caption{(a)  Time series of $\langle ({\nabla} J)^2 \rangle$ for the PS (black), 
SE (red), and FD (blue) runs 
plotted in this order for the $\eqr = 128$ case. PS and SE data still overlap, but FD the
FD result is noticeably different.
(b) Fractional error in $\Delta \langle ({\nabla} J)^2 \rangle$, 
as compared with a PS run with $2048 \times 2048$ resolution.
\label{gj2-128}}
\end{figure} 

\subsection{Accuracy with respect to higher order derivatives}
\label{sec_hoderivs}

In the above comparison of conservation laws, 
we see that the accuracy level can change with respect to field quantities involving 
a different order of spatial derivatives of $A$.
We now look into this issue further by comparing integrated field quantities.
Instead of presenting comparisons for all three levels of $\eqr$,
we will only present figures based on the $\eqr = 128$ case,
and just mention that similar conclusions can be drawn for the other two cases.

In Figs.~\ref{a2-128} to \ref{gj2-128}, we show in the (a) panels the time series of the quantities $\langle A^2 \rangle$,
$E = \frac{1}{2}\langle \b^2 + \u^2 \rangle$, $\langle J^2 \rangle$, and $\langle ({\nabla} J)^2 \rangle$,
for the PS run (black), SE run (red), and FD run (blue), 
plotted in this order for the $\eqr = 128$ case.
These four plots involve field quantities of increasing order of spatial derivatives of the 
magnetic potential $A$.
When plotting this way, it is not easy to see the difference between the curves when they are close to each other.
In fact, red curves almost totally cover black curves for all cases,
and blue curves almost totally cover the other two in $\langle A^2 \rangle$ and $E$.
So in the (b) panels, we plot the fractional difference $\Delta$ 
between these values and those from a ``converged'' PS run using
a much higher resolution with a uniform grid of $2048 \times 2048$.
Again, black is for the PS run (fractional difference between the $\eqr = 128$ PS run and the $2048$ PS run),
red is for the SE run, and blue is for the FD run.
The fractional difference (or error) $\Delta$ is defined as, e.g.,  $\Delta \langle A^2 \rangle \equiv \mid  \langle A^2 \rangle - \langle A^2 \rangle_{2048} \mid / \mid  \langle A^2 \rangle_{2048} \mid$. 

We note from the black curves that the PS run with a $128 \times 128$ grid is indeed a highly converged run,
in the sense that the fractional error with respect to the $2048 \times 2048$ is very small.
The error $\Delta \langle A^2 \rangle$ is at a level of about or slightly over $10^{-8}$.
This increases to around $5 \times 10^{-7}$ for $\Delta E$, 
around $10^{-6}$ to $10^{-5}$ for $\Delta \langle J^2 \rangle$,
and with a maximum value slightly below $10^{-3}$ for $\Delta \langle ({\nabla} J)^2 \rangle$.
The trend of obtaining less accurate results for quantities involving higher derivatives is expected.
Still, for all quantities that are important to the numerical integration, i.e. up to ${\nabla} J$, 
the accuracy of the $128 \times 128$ run is high enough for the whole duration of the simulation.
Again, this shows that the PS run can be used as a baseline for comparison.

For the SE run, the error level is higher than that of the PS run.
This is qualitatively similar to the comparison of conservation laws,
except now it is higher by a somewhat larger amount,
sometimes about two orders of magnitude.
Specifically, $\Delta \langle A^2 \rangle$ is at a level of about or a little over $10^{-5}$,
$\Delta E$  is up to about $5 \times 10^{-4}$, 
$\Delta \langle J^2 \rangle$ is about or a little over $10^{-3}$,
and $\Delta \langle ({\nabla} J)^2 \rangle$ is about or a little over $10^{-2}$ .
This level is still acceptably low, but is somewhat higher than those in
the error of conservation laws as shown above, 
especially for $\Delta \langle J^2 \rangle$ and $\Delta \langle ({\nabla} J)^2 \rangle$.
However, we need to keep in mind that the errors in the conservation laws are indicating
how well a code simulates the equation self-consistently at each moment,
but the errors in comparing with results from a converged solution are accumulated over time,
and can thus be larger than the former.

For the FD run, the error level is again higher than that of the SE run,
sometimes by an order of magnitude or more.
The error level of $\Delta \langle A^2 \rangle$ is about $10^{-3}$ or slightly above,
while $\Delta E$ is at a level or slightly over $10^{-2}$.
These two are still reasonably small and so we do not observe appreciable 
differences in their respective (a)-panel plots.
However, $\Delta \langle J^2 \rangle$ and $\Delta \langle ({\nabla} J)^2 \rangle$
become large enough to be observable in the time series plots themselves.
Quantitatively, $\Delta \langle J^2 \rangle$ is at a level of $10^{-1}$ or a slightly below,
while $\Delta \langle ({\nabla} J)^2 \rangle$ is at a level of $10^{-1}$ or somewhat above.
Although such high error levels do not seem to alter the overall dynamics of the
solutions qualitatively,
they are at a level high enough to be of some concern.

\subsection{Accuracy of current layer width}
\label{sec_currentwidth}

One quantity of great interest in the \ci\ problem is the width of current layers,
which can be defined as
\be
l \equiv  \left( \langle J^2 \rangle /  \langle ({\nabla} J)^2 \rangle \right)^{1/2} \, .
\label{l-def}
\ee

\begin{figure}
\plotone{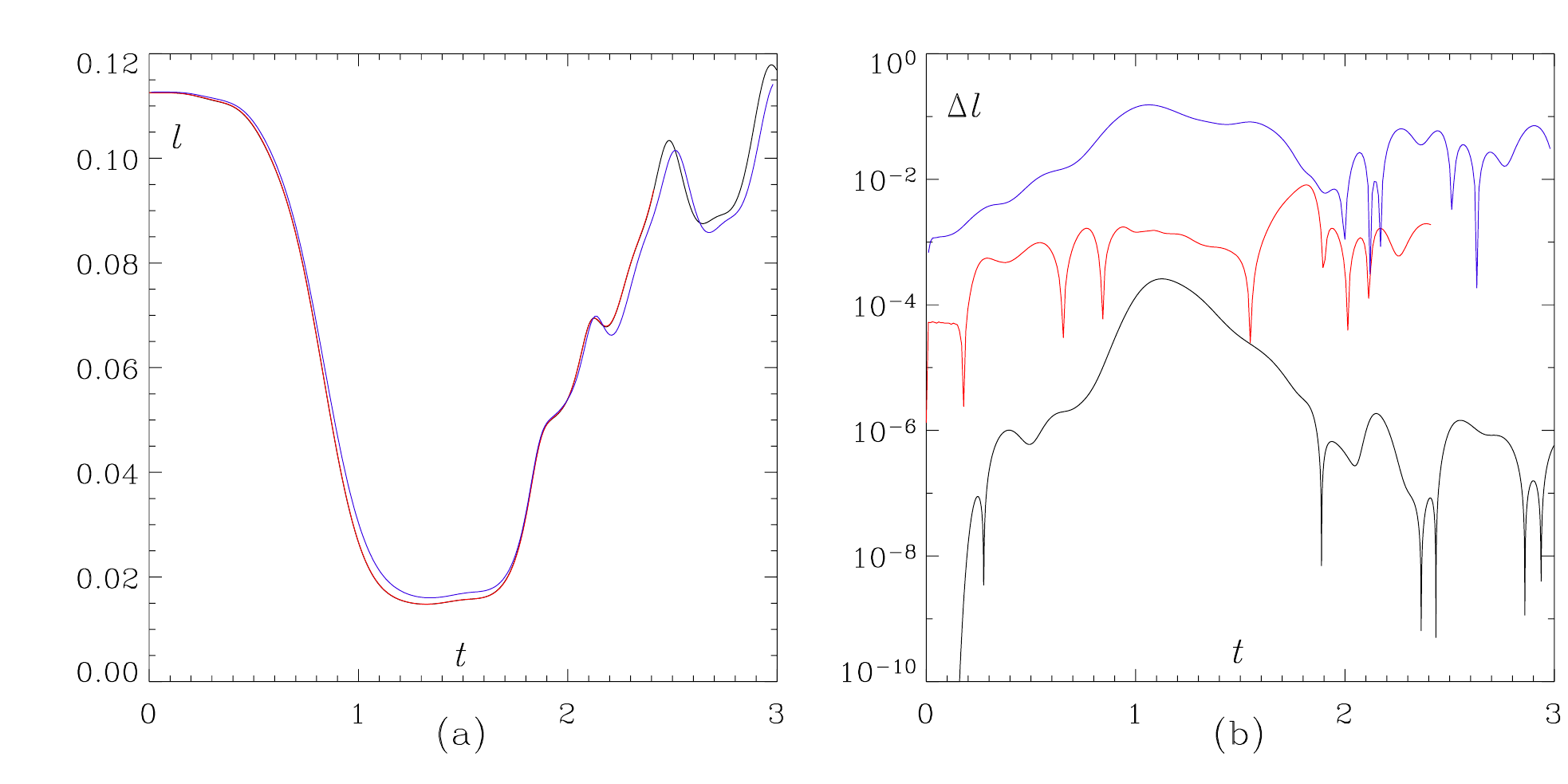}
\caption{(a)  Time series of the current layer width $l$ for the PS (black), 
SE (red), and FD (blue) runs 
plotted in this order for the $\eqr = 128$ case. 
(b) Fractional error in $\Delta l$, 
as compared with a PS run with $2048 \times 2048$ resolution.
\label{l-128}}
\end{figure}

In Fig.~\ref{l-128}, we show the current layer width $l$ defined in Eq.~(\ref{l-def}) for the PS (black), 
SE (red), and FD (blue) runs, 
plotted in this order for the $\eqr = 128$ case in panel (a),
and the error $\Delta l$, as compared with the $2048 \times 2048$ PS run in (b). 
We see that the error level $\Delta l$ is similar to that of $\Delta \langle ({\nabla} J)^2 \rangle$,
at about $10^{-1}$ or slightly over for the FD run, about an order of magnitude higher than that of the SE run.
Again, this accuracy level is qualitatively reasonable.
However, if we need to investigate the more difficult problem of the Parker's model (3D \ci)
and need to determine whether $l \rightarrow 0$ or not in the $\eta \rightarrow 0$ limit,
a 10\% error could lead to significant uncertainty.

\begin{figure}
\plotone{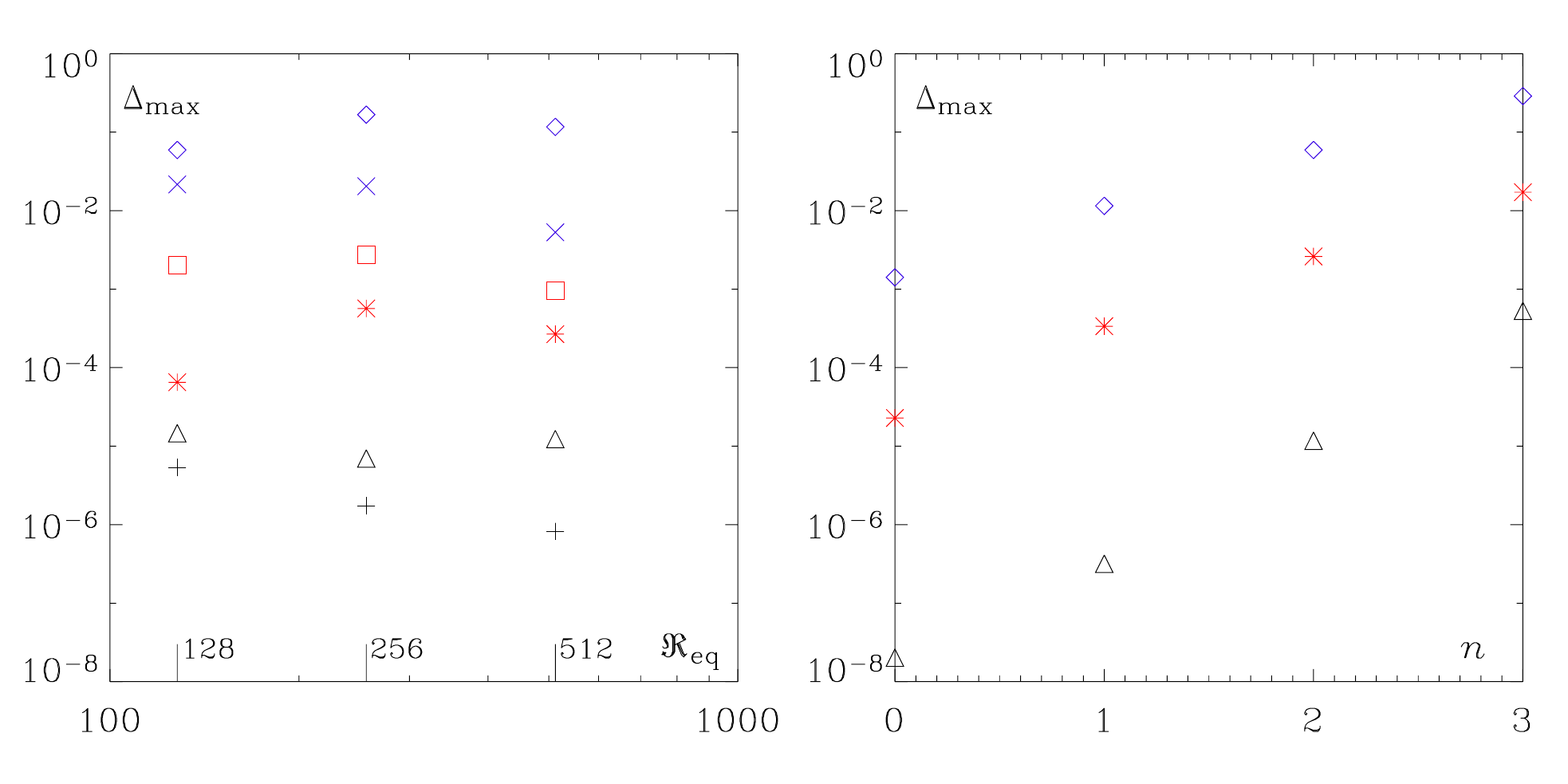}
\caption{(a)  Maximum fractional error over the plotted period in Fig.~\ref{c128} to
Fig.~\ref{c512} of  $\Delta \dot{E}$ (black triangles for PS, 
red asterisks for SE, blue diamonds for FD) and $\Delta \dot{M}$ (black plus signs for PS, 
red squares for SE, blue crosses for FD) for the three $\eqr$ cases, when compared
to a PS run at a grid resolution of $2048x\times2048$. 
(b) Maximum fractional error over the plotted period in Fig.~\ref{a2-128} to
Fig.~\ref{gj2-128} of  $\Delta \langle A^2 \rangle$ ($n = 0$),
$\Delta E$  ($n = 1$),  $\Delta \langle J^2 \rangle$ ($n = 2$), and
$\Delta \langle ({\nabla} J)^2 \rangle$ ($n = 3$) versus $n$, 
the order of spatial derivative with respect to $A$ (black triangles for PS, 
red asterisks for SE, blue diamonds for FD),
for the $\eqr = 128$ case.
\label{accu}}
\end{figure} 

\subsection{Quantitative summary}
\label{sec_errorsummary}

Let us summarize the comparisons from Fig.~\ref{c128}--~\ref{gj2-128} by plotting the maximum 
fractional error over the plotted time period for all cases.
In Fig.~\ref{accu}, we show in (a) the maximum fractional error over the plotted period in 
Fig.~\ref{c128} to Fig.~\ref{c512} of  $\Delta \dot{E}$ (black triangles for PS, 
red asterisks for SE, blue diamonds for FD) and
$\Delta \dot{M}$ (black plus signs for PS, 
red squares for SE, blue crosses for FD)
for the three $\eqr$ cases. 
In (b), we show the maximum fractional error over the plotted period in Fig.~\ref{a2-128} to
Fig.~\ref{gj2-128} of  $\Delta \langle A^2 \rangle$ ($n = 0$),
$\Delta E$  ($n = 1$),  $\Delta \langle J^2 \rangle$ ($n = 2$), and
$\Delta \langle ({\nabla} J)^2 \rangle$ ($n = 3$) versus $n$, 
the order of spatial derivative with respect to $A$ (black triangles for PS, 
red asterisks for SE, blue diamonds for FD),
for the $\eqr = 128$ case.
From these two plots, we see clearly the separation between the three groups
(black for PS, red for SE, and blue for FD).
While PS has the lowest level of error as expected (due to a fully spectral treatment
and to the fact that uniform grids are used),
SE achieves an error level somewhat in between PS and FD.
In some cases, the error in SE approaches that of PS, 
indicating near spectral accuracy.

\begin{figure}
\plotone{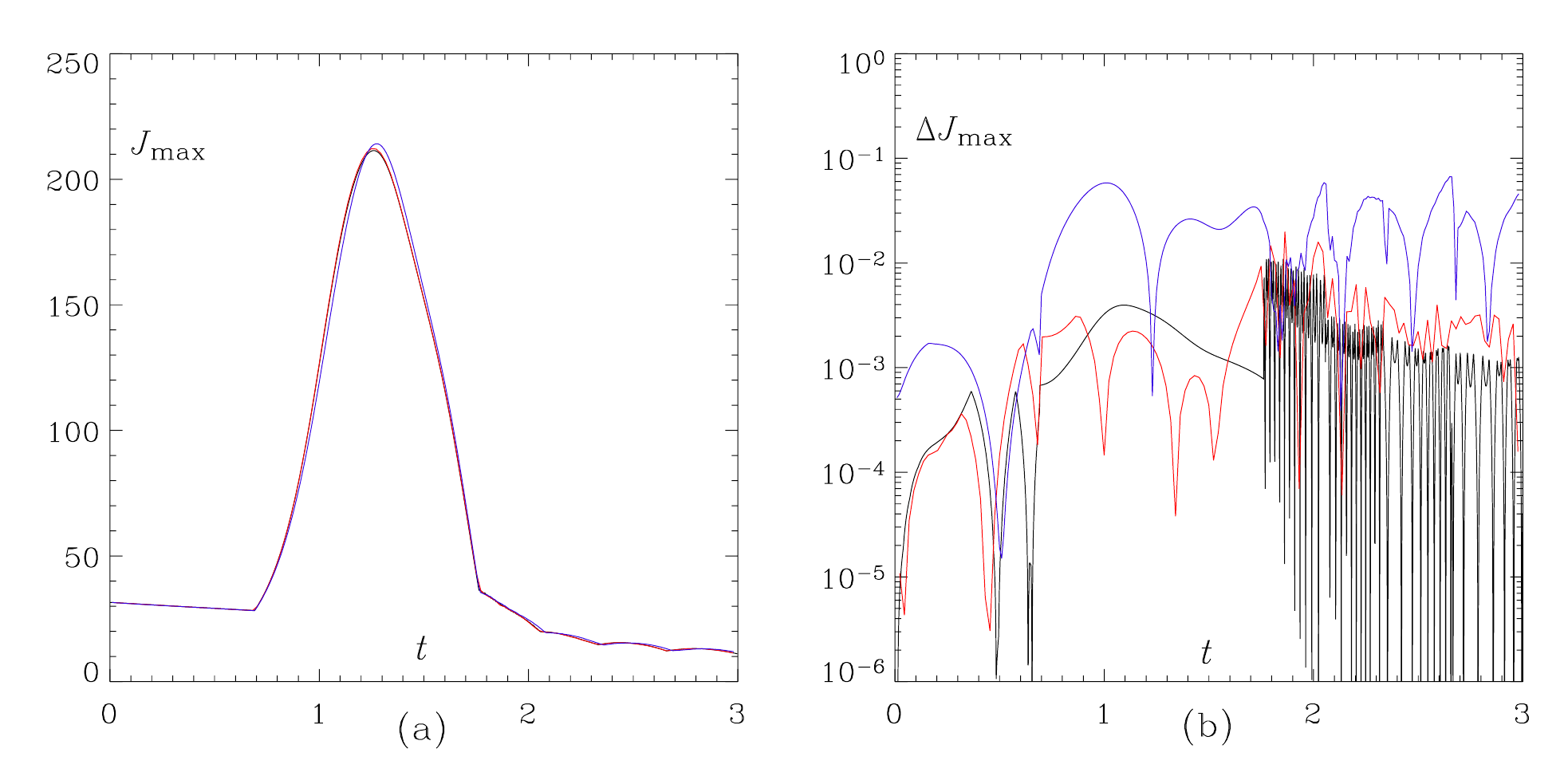}
\caption{(a)  Time series of $J_{\rm max}$ for the PS (black), 
SE (red), and FD (blue) runs plotted in this order for the $\eqr = 128$ case. 
(b) Fractional error in $\Delta J_{\rm max}$, 
as compared with a PS run with $2048 \times 2048$ resolution.
\label{jm-128}}
\end{figure} 

So far we have looked at comparisons of spatially integrated quantities. 
It is conceivable that the accuracy of field quantities at particular spatial points
may not follow similar trends.
Here we present one example that compares values of the maximum current density,
$J_{max}$, over the entire box. 
In Fig.~\ref{jm-128}, we show the maximum current density value over the periodic box,
$J_{\rm max}$, for the PS (black), SE (red), and FD (blue) runs, 
plotted in this order for the $\eqr = 128$ case in panel (a),
and the error $\Delta J_{\rm max}$, as compared with the $2048 \times 2048$ PS run in (b). 

We see in this comparison that even the PS run has an error level at around $10^{-2}$,
much higher than the errors of other quantities we have shown so far with this method.
This is because the grid used in the $2048 \times 2048$ run is much finer than what is used in the
$\eqr = 128$ PS run.
So, the value of $J_{\rm max}$ from a much higher resolution run cannot be located at the collocation
points of the lower resolution run, and thus it reaches a slightly higher value.
With this consideration,
the fact that both the PS and SE runs return 
$\Delta J_{\rm max}$ to about 1 \% accuracy is actually quite good.
At the same time, $\Delta J_{\rm max}$ of the FD run
is about an order of magnitude higher,  of about the same
level as $\Delta \langle J^2 \rangle$.

\begin{figure}
\plotone{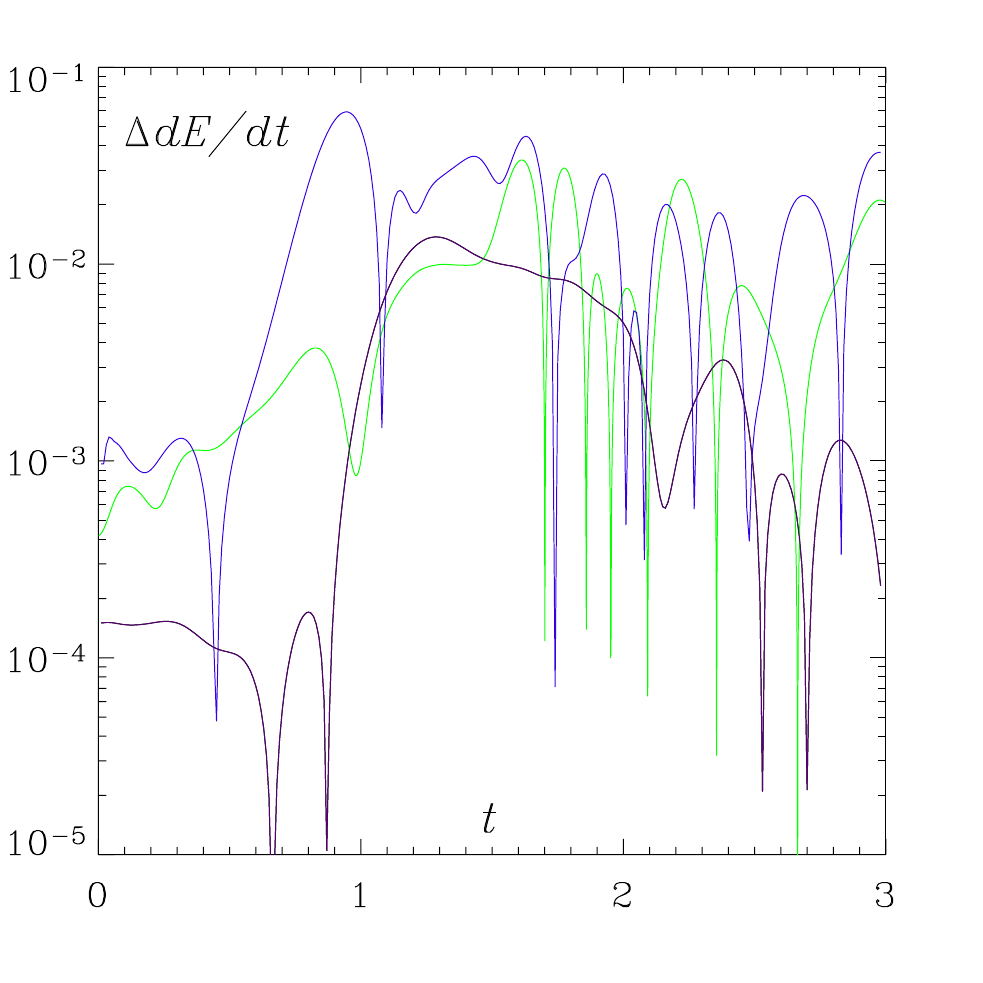}
\caption{The blue curve is the fractional error $\Delta \dot{E}$ of the FD run as shown in Fig.~\ref{c128}~(a).
The green curve is $\Delta \dot{E}$ of a FD run using the same grid as shown in 
Fig.~\ref{128-grids} but with
a $16 \times 16$ uniform grid within each square instead of $8 \times 8$ as used in the blue curve.
The purple curve is $\Delta \dot{E}$ of a FD run using a uniform grid of $256 \times 256$ resolution.
\label{fd-conv}}
\end{figure} 

\subsection{Comparison with FD at higher resolution}
\label{sec_hiresfd}

We must point out that in the above comparison,
we require the FD runs to use the same grid as the SE runs, with similar DOF.
This is done in order to obtain constrained comparisons that are easier to interpret. 
However, this will, of course, make the FD runs intrinsically less accurate,
as we have so far seen,
because FD schemes have a lower order truncation than spectral schemes.
In reality, one can compensate for this and increase the accuracy of FD runs by using higher resolution
or a greater number of DOF.
This will of course require more computational resources.
However, with FD schemes usually being more efficient (\eg, scalable), 
this may certainly be considered a reasonable way to obtain more accurate solutions.
Here we study briefly this possibility by running the FD code using higher resolutions.

In Fig.~\ref{fd-conv}, 
the blue curve is the fractional error $\Delta \dot{E}$ of the FD run as shown in Fig.~\ref{c128}~(a).
The green curve is $\Delta \dot{E}$ of a FD run using the same grid as shown in Fig.~\ref{128-grids} but with
a $16 \times 16$ uniform grid within each square instead of $8 \times 8$ as used in the blue curve.
The purple curve is $\Delta \dot{E}$ of a FD run using a uniform grid of $256 \times 256$ resolution.
We see that the green curve is mostly at a level below the blue curve, when the linear resolution is doubled.
The decrease is sometimes quite high, but at other times just marginal.
For the FD run using a $256 \times 256$ uniform grid, 
which has a cell size equals to the smallest cell size of run represented by the green curve, 
the error is substantially (about an order of magnitude) lower than that of the blue curve.
However, when compared with $\Delta \dot{E}$ of the SE run as shown in the red curve in Fig.~\ref{c128}~(a),
this is still about one to two orders of magnitude higher.

Of course, one can continue increasing the DOF of the FD run to try to reach the accuracy level 
of the SE run. Following the argument in \cite{rosenberg2007},
the linear error bound scaling for the SE case can be written as \cite[p. 273]{DFM2002}
$$
\epsilon_{SE}\sim h^{\min(p,s)} \ {p}^{-s} \ ,
$$
where $p$ is the polynomial expansion order,  $h\sim1/E$ is the uniform element length scale, and 
$s$ is the smoothness of the exact solution. For the moment we neglect prefactors in the
error term.
We assume $s=p$ so that derivatives can be computed up to the order of the 
method, suggesting a reasonably smooth function. The error for the 
FD method is
$$
\epsilon_{\rm FD}\sim h^{\beta} \sim N^{-\beta}  \ ,
$$
where $N$ is the total linear (equivalent) number of grid cells, and $\beta$ is the nominal
truncation order (typically $\approx 2$). Equating the logarithmic errors
yields 
$$
\log{N} = \frac{p}{\beta} \log (pE) + {\rm prefactor \; terms}.
$$
This scaling relation provides the number of FD cells required to achieve roughly the error
of the SE method at order $p$. We cannot actually use this scaling directly for comparisons
between the FD and SE methods, however, because the prefactor terms are unknown, and 
may depend critically on the flows. Still, we can illustrate the scaling for the case of the 
SE run in Fig.~\ref{c128}~(a). We have
that $E=16$ (the equivalently-resolved number of elements) and $p=8$, implying that 
$\log{N} \approx 16/\beta$, so with $\beta=2$, we find that achieving comparable 
accuracy in the FD method could lead to the requirement for a catastrophically large number of
cells, except, perhaps, in the case where the filling factor of the fine grid is extremely
small. This point will require further study. 

\section{Discussion and Conclusion} 
\label{sec_conclusion}

It is worth noting that at the present stage,
the spectral element method described in this paper is more 
costly in terms of computational time than the FD and PS methods 
with which we compare our results, the latter 
being optimal for periodic boundary conditions. 
We have not tried to compare CPU usage of the different codes due to the fact 
that the SE and FD codes are running on different computational platforms;
furthermore, both codes are still under continuous development, 
which can change efficiency greatly.
Instead, we concentrate on comparing accuracy obtained by the two schemes,
mainly to answer the question of whether using SE scheme has any advantage
that deserves putting more effort into further development.
We believe the results of this paper have shown that indeed the SE method
has substantial advantages and needs to be investigated further.

We have shown that using statically refined grids with DOF of roughly linear scaling,
SE method can produce results with high accuracy that can sometimes even
approach the spectral accuracy of the PS method.
This can be potentially very helpful in the investigation of the 3D \ci\ problem in the ideal limit,
which will require adaptive grids that can resolve distinct features.
The higher accuracy of the SE method can potentially provide a reliable definitive answer
to the important question of whether a true current sheet forms in the Parker's problem of
solar and stellar coronal heating.

Our main conclusion that SE methods can produce simulations with accuracy somewhat 
in between PS and FD methods is not surprising in itself.
However, our results yield important quantitative information in the context of
statically refined grids in problems with distinct spatial structures.

The accuracy of FD runs presented here is mainly for comparison purposes,
since we have imposed the restriction of using the same static grids with the same DOF.
This is by no means a suggestion that FD method cannot be used in the investigation of
the \ci, or similar problems.
Indeed we have also studied briefly that accuracy in FD method can also be increased
by using more DOF.
However, we have not studied the trade off of doing this with respect to the increase of the
usage of computational resources.
This may be an important topic for future investigation.

\acknowledgments

We acknowledge helpful discussions with Amik St.Cyr at NCAR.
Computer time was provided by NCAR and UNH. 
This research is supported in part by a NSF grant AST-0434322.
The NSF grant CMG-0327888 at NCAR also supported this work in part 
and is gratefully acknowledged.

\end{document}